

\magnification 1200
\hsize=14truecm
\hoffset=1truecm

\font\ttt=cmbx12  
\font\tt=cmbx10
\font\eightrm=cmr8
\font\ninerm=cmr9

\def\supp{\mathop{\rm supp}}
\def\dist{\mathop{\rm dist}}
\def\chapter{\bigskip\bigskip\noindent}
\def\echapter{\bigskip}
\def\section{\bigskip\noindent}
\def\esection{\medskip}


\begingroup
\font\thirteen=cmbx10 at 13pt
\font\ten=cmbx10
\font\eight=cmr8
\baselineskip 15pt
\nopagenumbers

\null
\bigskip\bigskip

\centerline{\ten Centre de Physique
Th\'eorique\footnote{$^{\star}$}{\eight Unit\'e Propre de 
Recherche 7061}, CNRS Luminy, Case 907}

\centerline{\ten F-13288 Marseille -- Cedex 9}

\vskip 2truecm

\centerline{\thirteen I N T E R F A C E S}

\bigskip

\centerline{{\bf 
Salvador MIRACLE-SOLE\footnote{$^1$}{\eight 
Centre de Physique Th\'eorique, CNRS, Marseille}
}}

\vskip 2truecm

\centerline{\bf Abstract} 

\medskip

Several aspects of the theory of the coexistence of phases 
and equilibrium forms are discussed.
In section 1, the problem is studied from the point of view 
of thermodynamics. 
In section 2, the statistical mechanical theory is 
introduced. 
We consider, in particular, the description of the microscopic 
interface at low temperatures and the existence of a free 
energy per unit area (surface tension). 
In section 3, a proof is given of the microscopic validity 
of the Wulff construction in a simplified model of a 
two-dimensional interface. 
Finally, in section 4, the roughening transition and 
the formation of facets in an equilibrium crystal are 
studied. 
Appendices A and B concern, respectively, the first and 
second points of section 2, mentioned above.


\bigskip

Published in 
{ \it Cellular Automata and Complex Systems,}
E. Goles and S. Martinez (eds.), 
Kluwer, Dordrecht, 1999, pp.\ 107--151. 
Proceedings of the V Fiesta (F{\'\i}sica Estad{\'\i}stica) 
Conference, December 9--13, 1996, Santiago, Chile.

\bigskip

\smallskip



\bigskip

\noindent CPT-98/P. 3651

\bigskip



\vfill\eject
\endgroup


\headline{\hfill\sl S. Miracle-Sole / 
Interfaces\hfill}

\noindent
{\ttt 1\ \  Introduction}
\echapter

When a fluid is in contact with another fluid, or with a gas, 
a portion of the total free energy of the system is 
proportional to the area of the surface of contact, 
it is also proportional to a coefficient, the surface tension,
which is specific for each pair of substances. 
Equilibrium, which is the condition of minimum free 
energy in the system, will accordingly be obtained 
by the utmost possible reduction of the surfaces in contact. 

We must understand here, not an absolute but a relative 
minimum, depending on the material circumstances of the case 
that is considered.
Mathematicians restrict the term minimal surfaces to the
former cases, or more generally, to all cases in which
the mean curvature is zero. Other surfaces which are only 
minimal with respect to the volume contained, are called
surfaces of constant mean curvature. From the fact that we 
may extend a soap film across any ring of wire, 
however the wire be bent, we see that there
is no end to the variety of the minimal surfaces that can
be constructed or imagined. To fit a minimal surface to
the boundary of any given closed curve in the space is
a problem formulated by Lagrange, commonly known
as the problem of Plateau, who solved it with the 
soap films.

When we have three fluids in contact with one another, new
interesting situations arise. Suppose that a drop of some
fluid, $a$, float on another fluid, $b$, while both are exposed
to air, $c$. Then we have three different surfaces of contact,
and the total free energy of the system consists of three
parts, associated to these three surfaces. 
Equilibrium will be reached by contracting the surfaces
whose surface tension happens to be large and extending
those where it is small. 
A drop of fluid $a$, will exist
provided its own two surface tensions exceed the surface
tension between fluid $b$ and the air, i.e., provided that
$$
\tau_{ab}+\tau_{ac}>\tau_{bc}
$$
If equality is attained, then a film of fluid $a$ 
is formed, a situation which is known as perfect wetting.
For instance, floating on water, turpentine gathers into
a drop and olive oil spreads out in a film.

When one of the substances involved is anisotropic, such
as a crystal, the equilibrium is not determined as it was
in the case of a fluid, by the condition of minimal area.
The contribution to the total free energy of each element
of area of the crystal surface depends on its orientation
with respect to the crystalline axes. 

Gibbs' article ``On the equilibrium of heterogeneous 
substances'' (Gibbs 1875)
is a theoretical system of thermodynamics, of nearly 300 pages,
derived from the first and second law. In this work, Gibbs
introduced the principle of minimum of total surface free
energy. He also showed the role of the anisotropic surface
tension for determination of the shape of a crystal in
equilibrium. 
Concerning the formation of facets, he remarks: 

\medskip
\itemitem{}
``the value of $\tau$ varies greatly with the direction of the
surface with respect to the axes of crystallization and
in such a manner as to have sharply defined minima
(The differential coefficients of $\tau$ with respect to the
direction-cosines of the surface appear to be discontinuous
functions of the latter quantities).''
\medskip

\noindent
He also points out the complexities of actual growth of a
crystal and suggests that only small crystals have an ideal
equilibrium form:

\medskip
\itemitem{}
``it seems not improbable that the form of very minute
crystals in equilibrium is principally determined by the
condition that the free energy shall be a minimum for the
volume of the crystal, but as they grow larger the deposition
of new matter on the different surfaces will be determined
more by the nature (orientation) of the surface and less
by their size in relation to the surrounding surfaces.
As a final result, a large crystal thus formed will 
generally be bounded by those surfaces alone on which the
deposition of new matter takes place least readily''
\medskip

Curie (1885) considered also, independently of Gibbs
and using a different method, 
a crystal in equilibrium with its own vapor phase, 
arriving at the same conclusion. 
For a crystal of $N$ faces, of area $F_i$ and specific
surface free energies $\tau_i$, he found that 
$$
f=\tau_1 F_1+\tau_2 F_2+\dots+\tau_N F_N
$$
has the minimum value for a given volume. 
As an application he examined two problems,  
a crystal taking the form of a parallelepiped, and  
a crystal taking the form of a of a cubo-octahedron. 
The second problem, for which Curie gives only the 
result, is not obvious.

The article by Curie was the point of departure for Wulff's 
classical experiments on crystal growth. 
He reported the results in a paper (Wulff 1901), 
first published in Russian in 1895. 
His principal conclusions include the celebrated 
Wulff's theorem on the shape of the equilibrium crystal:

\medskip
\itemitem{}
``The minimum surface energy for a given volume of a 
polyhedron will be achieved if the distances of its faces 
from one and the same point are proportional to their 
capillary constants.'' 
\medskip

\noindent
The term capillary constants was used for the surface tension. 
Denoting these distances by $h_i$, this means that 
$$
{h_1\over\tau_1}={h_2\over\tau_2}=\cdots={h_N\over\tau_N}
$$
Wulff himself supported his principle mainly by its  
consequences (relating the rate of growth in various 
directions to the corresponding surface tensions). 
His attempt at a general proof failed. 
He surmised that the surface free energy $f$ 
(given by Curie's formula) and the volume $V$ 
are related to $h_i$ by the following expressions 
$$
f=p\sum\tau_ih_i^2, \quad  V=q\sum h_i^3
$$
where $p$ and $q$ are constants. 
However, the simplest example of a parallelepiped contradicts
these statements.

The complete proof of Wulff theorem was presented 
by von Laue (1943), Herring (1951), and others. 

The surfaces tensions are dependent upon the geometric
distribution of the particles making up the crystal
structure. 
In this way the Wulff theorem establish a relation
between the forms and the structure of crystals. 
Although it is not easy to find equilibrium 
crystals in nature we may think, following Schneer (1970), 
that the ensemble of the forms of a mineral species
should approach equilibrium. 
The distribution of these forms in
geological time and space is the outcome of 
a world-experiment establishing the probability of
occurrence of each particular form. 

A complete map of the mineral kingdom known to science 
was presented in Goldschmidt Atlas (1913--1923). 
Each of the nine volumes of illustrations was accompanied
by a volume of morphological data. 
Ironically, the first volume of the Atlas appeared in 1913, 
one year after the discovery of X-ray difraction. 
Its purpose was to provide data for the determination of 
the molecular forces within the crystal from the crystal
form. 

It is only in recent times that equilibrium crystals 
have been produced in the laboratory. 
Most crystals grow under non-equilibrium conditions, 
as predicted by Gibbs, and is a subsequent relaxation 
of the macroscopic crystal that restores the equilibrium. 
This requires transport of material over long distances 
and the time scales can be very long, even for very small 
crystals. 
On has been able to study, however, metal crystals 
in equilibrium of size 1--10 micron (Heyraud and 
Metois 1980, 1983). Equilibration times of a few days 
were observed. 

A very interesting phenomenon that can be observed on 
equilibrium crystals is the roughening transition.
This transition is characterized by the disappearance of 
a facet of a given orientation from the equilibrium
crystal, when the temperature attains a certain 
particular value. 
Roughening transitions have been found experimentally,  
first, in negative crystals, i.e., vapor bubbles included in a
crystal, in organic substances 
(Pavlovsca and Nenov 1971, 1973). 
The best observations have been made on helium crystals in 
equilibrium with superfluid helium 
(Balibar and Castaign 1980, 1985, 
Keshishev et al.\ 1981, Wolf et al.\ 1983),   
since the transport of matter and heat is extremely fast 
in the superfluid. 
Crystals grow to size of 1--5 mm and relaxation times vary 
from milliseconds to minutes. 
Roughening transitions for three different types of facets 
have been observed.

\vfill\eject

\chapter
{\ttt 1\ \ Thermodynamics of equilibrium forms}
\echapter

\section
{\tt 1.1\ \ Variational problem}
\esection

Our aim is to find the shape of a droplet of a phase $c$, 
the crystal, inside a phase $m$, called the medium,
when both phases are in equilibrium.
The phase $c$ need not be a crystal, this terminology
is adopted to stress that we consider that $c$ has
anisotropic properties.
This shape is obtained by minimizing the 
surface free energy between the phases $c$ and $m$.

Let ${\bf n}$ be a unit vector in ${\bf R}^d$ and 
consider the situation in which the phases $c$ and $m$
coexist over a hyperplane perpendicular to ${\bf n}$, 
we denote by $\tau({\bf n})$ the surface tension,
or free energy per unit area, of such an interface. 
We consider the surface tension $\tau({\bf n})$ as
a positive function on the unit sphere 
${\bf S}^{d-1}\subset{\bf R}^d$, but we do not require
now the symmetry $\tau({\bf n})=\tau(-{\bf n})$.
In the case under consideration $\tau({\bf n})$ is 
obviously a symmetric function, but 
non-symmetric situations are also physically 
interesting and they appear, for instance, in the case of 
a sessile drop on a wall to be be discussed in
section 1.6. 

We denote by $B$ the set of ${\bf R}^d$ occupied
by the phase $c$, and by $\partial B$ the boundary of $B$.
The total surface free energy of the crystal is 
given by
$$
\tau(\partial B)=\int_{\xi\in\partial B}\tau({\bf n}(\xi))
ds_\xi
\eqno(1.1)$$
Here ${\bf n}(\xi)$ is the exterior unit normal to
$\partial B$ at $\xi$ and $ds_\xi$ is the element of area at
this point. 
We suppose that the boundary of $B$ is sufficiently
smooth so that the normal is defined for almost all
$s$ in $\partial B$.

The variational problem is to minimize (1.1) under the 
constraint that the total (Lebesgue) volume $\vert B\vert$ 
occupied by the phase $c$ is fixed. 
Given a set ${\cal W}$, we say that the crystal $B$ has 
shape ${\cal W}$ if after a translation and a dilation it
equals ${\cal W}$.
The solution of the variational problem, known under
the name of Wulff construction, is given below.
Notice that the
problem is scale invariant, so that if we can solve
it for a given volume of the phase $c$, we get the 
solution for other volumes by an appropriate scaling.
Let ${\cal W}$ be defined by 
$$
{\cal W}=\{{\bf x}\in{\bf R}^d : 
{\bf x}\cdot{\bf n}\le\tau({\bf n})\}
\eqno(1.2)$$
where the inequality is assumed for every 
${\bf n}\in{\bf S}^{d-1}$
(${\bf x}\cdot{\bf n}$ denotes the euclidean scalar product).
We shall show that this set, which will be called the 
Wulff shape, gives the optimal shape for the crystal.

\medskip

{\tt Theorem 1.1. } {\it 
Let ${\cal W}$ be the Wulff shape (1.2) for the surface 
tension function $\tau({\bf n})$.
Let $B\subset{\bf R}^d$ be any other region with 
sufficiently smooth 
boundary and the same (Lebesgue) volume as ${\cal W}$.
Then 
$$
\tau(\partial B)\ge\tau(\partial{\cal W})
\eqno(1.3)$$
with equality if and only if $B$ and ${\cal W}$ have the 
same shape.}

\medskip

A proof of this Theorem, based on some version of 
the isoperimetrical inequality,
will be given at the end of section 2.2 
(in this proof we follow Taylor 1987).

We now state some elementary properties of ${\cal W}$.
Being defined in (1.2) as the intersection of closed 
half-spaces
the Wulff shape ${\cal W}$ is a closed, bounded convex set, 
i.e., a convex body. 

Among the functions $\tau({\bf n})$, which through (1.2)
define the same shape ${\cal W}$,
there is a unique function having the property
that all planes 
$\{{\bf x}\in{\bf R}^d:{\bf x}\cdot{\bf n}=
\tau({\bf n})\}$, 
associated to all different unit vectors ${\bf n}$,
are tangent to the convex set ${\cal W}$.
This function is given by
$$
\tau_{\cal W}({\bf n})=
\sup_{{\bf x}\in{\cal W}}({\bf x}\cdot{\bf n})
\eqno(1.4)$$
and is called the support function of the convex body 
${\cal W}$.
If we consider an arbitrary function $\tau({\bf n})$
defining the Wulff shape ${\cal W}$,
it can be that some of these planes do not
touch the set ${\cal W}$. 
In other words, we have always 
$$
\tau({\bf n})\ge\tau_{\cal W}({\bf n})
\eqno(1.5)$$
but the inequality sign can hold for some ${\bf n}$.
The support function $\tau_{\cal W}({\bf n})$ is 
the smallest function on the unit sphere
which gives the Wulff shape ${\cal W}$. 

\section
{\tt 1.2\ \  Geometric inequalities}
\esection

We can state as follows the isoperimetric property 
of convex bodies.

\medskip
{\tt Theorem 1.2. } {\it 
Let ${\cal W}\subset{\bf R}^d$ be a convex body and 
$\tau_{\cal W}({\bf n})$ the corresponding support 
function. 
For any set $B\subset{\bf R}^d$, with sufficiently 
smooth boundary, we have 
$$
\tau_{\cal W}(\partial B)\ge d\ 
\vert{\cal W}\vert^{1/d}\vert B\vert^{(d-1)/d}
\eqno(1.6)$$
where $\vert{\cal W}\vert$, $\vert B\vert$, denote the
(Lebesgue) volumes of ${\cal W}$, $B$, respectively,
and $\tau_{\cal W}(\partial B)$ is the surface
functional defined by (1.1).
The equality occurs only when $B$ and ${\cal W}$
have the same shape.}
\medskip

Let us recall the following classical statement, 
known as the isoperimetric property of the circle.

\medskip

{\tt Proposition. }
The area $F$ and the length $L$ of any plane domain with
rectifiable boundary satisfy the inequality
$$
L^2\ge 4\pi F
\eqno(1.7)$$
The equality sign holds only for the circle.

\medskip

If for the set ${\cal W}$ we choose $D$, the closed circle 
of unit radius with center at the origin, then the
corresponding support function $\tau_{\cal D}({\bf n})$ 
is equal to the constant 1, and we obtain the
isoperimetric property. 

Mathematicians were convinced that the isoperimetric 
property of the circle is valid since ancient times. 
The known proofs of inequality (1.7) may be grouped
together around several main approaches.
Three of them go back to Steiner, who gave
the first non rigorous proofs of the isoperimetric 
property of the circle on the plane and of the
sphere and proved a similar property of the ball
in space.

We shall describe here another approach based on the 
notion of vector addition of sets due to Minkowski.
This is a very general approach which easily leads
to the Brunn-Minkowski inequality and, as
a particular case, to the isoperimetric inequality. 

To every pair of non-empty sets $A,B\subset{\bf R}^d$
their vector Minkowski sum is defined by
$A+B=\{a+b : a\in A,b\in B\}$. 
If $A,B$ are compact sets, then $A+B$ is compact.
In this case each of the sets necessarily has a volume, 
its Lebesgue measure. 
The following Theorem is known as the
Brunn-Minkowski inequality.

\medskip

{\tt Theorem 1.3. } {\it 
For non empty compact sets $A,B\subset{\bf R}^d$
$$
\vert A+B\vert^{1/d}\ge
\vert A\vert^{1/d}+\vert B\vert^{1/d}
\eqno(1.8)$$
and the equality sign in (1.8) holds only in the following
cases: $\vert A+B\vert=0$, one of the sets $A,B$ consists
of a single point, and $A$ and $B$ are two convex bodies 
with the same shape.}

\medskip

A proof of this Theorem can be found in the
book by Burago and Zalgaller (1988).
First one proves by direct computation that the inequality 
holds in the case in which $A$ and $B$ are parallelepipeds 
with sides parallel to the coordinate axis.
The validity of the inequality is then extended by 
induction to all finite unions of such parallelepipeds and,  
finally, to all compact sets by an appropriate limit 
process.

The following geometrical remark will next be used to 
prove the isoperimetric inequality. 

\medskip

{\it Remark. }
Let ${\cal W}$ be a convex body in ${\bf R}^d$. 
Given any set $B\subset{\bf R}^d$, with sufficiently 
smooth boundary, we can express
the functional $\tau_{\cal W}(\partial B)$ as
$$
\tau_{\cal W}(\partial B)
=\lim_{\lambda\to0}{{\vert B+\lambda{\cal W}\vert
-\vert B\vert}\over\lambda}
\eqno(1.9)$$
where $\lambda{\cal W}$ denotes the homothetic set 
$\{\lambda{\bf x}:{\bf x}\in{\cal W}\}$. 
If $B\subset{\bf R}^d$ is bounded, the functional
$\tau_{\cal W}$ is well defined.
In particular, (1.9) shows that 
$$
\tau_{\cal W}(\partial{\cal W})=d\,\vert{\cal W}\vert
\eqno(1.10)$$

\medskip

{\it Proof of Theorem 1.2. }
We first prove equation (1.9).
In fact, if $H({\bf n})$ denotes the half space whose 
oriented unit normal is ${\bf n}$, we obtain
from the definition of $\tau_{\cal W}$
$$
\tau_{\cal W}({\bf n})=\dist\big(\partial H({\bf n}),
\partial(H({\bf n})+{\cal W})\big)
\eqno(1.11)$$
We may then write
$$\eqalignno{
\tau_{\cal W}(\partial B)
&=\int_{\xi\in\partial B}\dist\big(\partial H({\bf n}(\xi)),
\partial(H({\bf n}(\xi))+{\cal W})\big)ds_\xi \cr
&=\int_{\xi\in\partial B}\lim_{\lambda\to0}(1/\lambda)
\dist\big(\xi,\partial(B+\lambda{\cal W})\big)ds_\xi 
&(1.12)\cr}$$
This last expression coincides with the right hand 
side of (1.9), and proves the validity of this 
equation.

Then inequality (1.6) follows by applying the 
Brunn-Minkowski inequality to 
$\vert B+\lambda{\cal W}\vert$.
This gives
$$
\vert B+\lambda{\cal W}\vert-\vert B\vert\ge
\big(\vert B\vert^{1/d}+\lambda\vert{\cal W}\vert^{1/d}
\big)^d-\vert B\vert\ge
d\,\lambda\,\vert B\vert^{(d-1)/d}\vert{\cal W}\vert^{1/d}
\eqno(1.13)$$
which, taking equation (1.9) into account, ends the
proof of inequality (1.6).

By taking limits we can deduce (1.6) from (1.8).
However, conditions for the equality to occur in (1.6)
cannot be found in this manner.
For a proof that ${\cal W}$ is the unique minimum
see Burago and Zalgaller (1988).

\medskip

{\it Proof of Theorem 1.1. }
The proof can be obtained as a consequence of the 
Theorem 1.2. 
Let ${\cal W}$ be the Wulff shape corresponding to 
the function $\tau$.
Then
$$
\tau(\partial B)\ge\tau_{\cal W}(\partial B) 
\ge d\ \vert{\cal W}\vert^{1/d}\vert B\vert^{(d-1)/d} 
\eqno(1.14)$$
taking into account equation (1.9) and 
the isoperimetric inequality (1.6).
But, when $B={\cal W}$, we have
$$
\tau(\partial{\cal W})=
\tau_{\cal W}(\partial{\cal W})=d\ \vert{\cal W}\vert
\eqno(1.15)$$
The first equality in (1.15) follows from the fact that 
$\tau_{\cal W}({\bf n})\ne\tau({\bf n})$
only for the unit vectors ${\bf n}$ for which 
the planes ${\bf x}\cdot{\bf n}=\tau({\bf n})$
are not tangent to the convex set ${\cal W}$.
The second equality follows from equation (1.9).
Therefore
$$
\tau(\partial  B)\ge \tau(\partial{\cal W})\,
(\vert B\vert/\vert{\cal W}\vert)^{(d-1)/d}
\eqno(1.16)$$
which, when $\vert B\vert=\vert{\cal W}\vert$,
gives the stated inequality (1.3).
The equality in (1.3) corresponds to the equality in (1.6).
The Theorem is proved.

\vfill\eject

\section
{\tt 1.3\ \ Convexity properties}
\esection

It is expected, from thermodynamic reasons, that the 
surface tension $\tau({\bf n})$ is equal to the support 
function of the associated Wulff shape ${\cal W}$.
We next comment on this point, 
first discussed by Dobrushin and Shlosman (1992) 
and Messager et al.\ (1992).

Let $A_0,\dots,A_d\in{\bf R}^d$ 
be any set of $d+1$ points in general position and, for
$i=0,\dots,d$, let $\Delta_i$ 
be the $(d-1)$-dimensional simplex defined by all these 
points except $A_i$. 
Denote by $\vert\Delta_i\vert$ 
the $(d-1)$-dimensional area of $\Delta_i$ and by
${\bf n}_i$ the unit vector orthogonal to $\Delta_i$.
The first vector, ${\bf n}_0$, is oriented toward the 
exterior of the simplex $A_0,\dots,A_d$, while the others,
${\bf n}_i$, $i=1,\dots,d$, are oriented inside.
We say that $\tau({\bf n})$ 
satisfies the pyramidal inequality if 
$$
\vert\Delta_0\vert \, \tau({\bf n}_0)
\le \sum^d_{i=1}\vert\Delta_i\vert\, \tau({\bf n}_i)
\eqno(1.17)$$
for any set $A_0,\dots ,A_d$.
We introduce the function on ${\bf R}^d$ defined by
$$
f({\bf x})=\vert{\bf x}\vert\, 
\tau({\bf x}/\vert{\bf x}\vert)  
\eqno(1.18)$$
If this function is convex, it satisfies
$$ \eqalignno{
f(\alpha {\bf x}) &= \alpha f({\bf x}) &(1.19) \cr
f({\bf x}+{\bf y}) &\le f({\bf x})+f({\bf y}) &(1.20) \cr
}$$
for any $\alpha >0$ and any ${\bf x}$ and ${\bf y}$ in 
${\bf R}^d$.
The first condition, 
that $f({\bf x})$ is a 
positively homogeneous convex function of degree 1,
is clear from the definition.
The second is equivalent in this case to the 
convexity condition.

\medskip

{\tt Theorem 1.4. } {\it
The following propositions are equivalent
\item{1)}
$\tau({\bf n})$ satisfies the pyramidal inequality   
\item{2)}
the function $f({\bf x})$  
is convex
\item{3)}
$\tau({\bf n})$ is the support function of the 
convex body ${\cal W}$ }

\medskip

{\it Proof. }
For simplicity we shall restrict ourselves to the
three dimensional case. 
Assume that condition 3) is satisfied and let
us consider the set ${\cal W}$ and its tangent
planes orthogonal to ${\bf n}_0,\dots,{\bf n}_3$.
These planes form a pyramid whose four vertices are
denoted by $A_0,A_1,A_2,A_3$. 
The notation is chosen in such a way
that the plane $A_1,A_2,A_3$ separates the origin $O$
from the vertex $A_0$.
This is possible, for all such pyramids, 
if and only if ${\cal W}$ is convex. 
We consider also the four pyramids having as
vertices the origin and three of these points. 
The volumes of these pyramids satisfy
$$
\vert OA_1A_2A_3\vert\le\vert OA_0A_2A_3\vert+
\vert OA_0A_3A_1\vert+\vert OA_0A_1A_2\vert
\eqno(1.21)$$
Taking into account that the heights of these pyramids
are $\tau({\bf n}_0)$, $\tau({\bf n}_1)$, $\tau({\bf n}_2)$
and $\tau({\bf n}_3)$, respectively, we get the 
pyramidal inequality for the pyramid $A_0,A_1,A_2,A_3$.
This shows that conditions 1) and 3) are equivalent.

Next, we observe that
$$
\vert\Delta_0\vert\,{\bf n}_0
=\vert\Delta_1\vert\,{\bf n}_1+\vert\Delta_2\vert\, 
{\bf n}_2 + \vert\Delta_3\vert\,{\bf n}_3
\eqno(1.22)$$
Thus the pyramidal inequality says that
$$
f({\bf x}_1+{\bf x}_2+{\bf x}_3)\leq 
f({\bf x}_1)+f({\bf x}_2)+f({\bf x}_3)
\eqno(1.23)$$
with 
${\bf x}_k=\vert\Delta_k\vert\,{\bf n}_k,\ k=1,\dots,3$. 
Then, to prove (1.20), it is enough to find pyramids 
such that 
$$
\vert\Delta_1\vert\,{\bf n}_1\rightarrow{\bf x},\quad
\vert\Delta_2\vert\,{\bf n}_2\rightarrow{\bf y},\quad 
\vert\Delta_3\vert\,{\bf n}_3\rightarrow 0 
\eqno (1.24)$$
for all ${\bf x}$ and ${\bf y}$. 
This can easily be done and proves the equivalence
of 1) and 2).

\medskip

The pyramidal inequality may be interpreted as a 
thermodynamic stability condition and thus also the 
convexity of $f({\bf x})$. 
If one supposes that 
$ \vert\Delta_0\vert \, \tau({\bf n}_0) $
is greater than the right hand side of inequality (1.17)
this would make the interface bounded by the sides
of $\Delta _0$ unstable and difficult to realize.

Let us mention that the term of support function 
is usually attributed to the convex function $f({\bf x})$. 
In fact a support function can be defined,  
by the same expression, for any convex set in 
${\bf R}^d$, and it is a real (or $+\infty$) valued 
function, positively homogeneous and convex.   
For a non-empty convex body 
the convex function $f({\bf x})$ is finite and hence 
(see e.g. Ruelle 1969, Proposition 3.3.4) 
Lipschitz continuous.

If $f({\bf x})=f(-{\bf x})$ the convex body 
${\cal W}$ is symmetric with respect to the origin. 
Now, if ${\cal W}$ is a symmetric closed bounded convex set 
and the origin belongs to the interior of ${\cal W}$, then 
the support function of ${\cal W}$ is finite everywhere, 
symmetric and strictly positive except at the origin, 
i.e., it is a norm. 
Let then  
${\tilde f}({\bf x}) = \sup_{\bf y}\big({\bf x}
\cdot {\bf y}/f({\bf y})\big)$ 
be the dual norm. 
In this case, the Wulff equilibrium shape ${\cal W}$
can be interpreted as the unit ball
with respect to the dual norm 
$$
{\cal W} =\{{\bf x}\in{\bf R}^d : 
{\tilde f}({\bf x})\le 1\} 
\eqno(1.25)$$ 


\section
{\tt 1.4\ \ Legendre transformation}
\esection

The Wulff construction can also be viewed as a 
geometrical version of the Legendre transformation.
We consider the function $f({\bf x})$, defined  
from the surface tension $\tau({\bf n})$ 
by equation (1.18), but we do not require now that $f$
is the support function of a convex body.
From definition (1.2), we get
$$
{\cal W} =\{{\bf x}\in{\bf R}^d : f^*({\bf x})\le0\}
\eqno(1.26)$$ 
where $f^*$ is the Legendre transform of $f$,
$$
f^*({\bf x}) = \sup_{\bf y}\big({\bf x}\cdot{\bf y}
-f({\bf y})\big)
\eqno(1.27)$$
Actually $f^*({\bf x})=0$ if ${\bf x}\in{\cal W}$,
and $f^*({\bf x})=\infty$ otherwise.
The support function of ${\cal W}$ is given by
$f^{**}$, the Legendre transform of $f^*$.

This point of view was developed by Andreev (1981).
For simplicity we consider the case $d=3$ and 
assume also that ${\cal W}$ is symmetric with  
respect to the origin.
We introduce a function $\varphi$ on ${\bf R}^2$ 
such that the graph of
$x_3 = \varphi(x_1,x_2)$, for $x_3>0$,
coincides with the boundary  $\partial{\cal W}$
of the crystal shape.
Since ${\cal W}$ is a convex body, $\varphi$ is a 
concave function, and
$$ 
{\cal W}=\{{\bf x}\in{\bf R}^3 : 
-\varphi(-x_1,-x_2) \le x_3 \leq \varphi (x_1,x_2)\} 
\eqno(1.28)$$ 
In the present context this means that $-\varphi$
is the Legendre transform of the
projected surface tension $\tau_p=(1/n_3)\tau$, 
considered as a function on ${\bf R}^2$
of the slopes $x_1=n_1/n_3,x_2=n_2/n_3)$.  
In other words,
$$ 
\tau_p (x_1,x_2) = f(x_1,x_2,1) 
\eqno (1.29)$$ 
Indeed, from equations (1.26) and (1.27), we see that 
$$ 
-\varphi(x_1,x_2) =
\sup_{y_1,y_2}\big(x_1y_1+x_2y_2-\tau_p(y_1,y_2)\big) 
\eqno(1.30)$$

Formula (1.28) is known as the Andreev construction.
The interest of this approach comes from the fact
that $\varphi$, and hence, 
the crystal shape itself, may be regarded as the free 
energy associated to a certain statistical mechanical 
Gibbs ensemble. 
We shall consider this ensemble in section 3.

\section
{\tt 1.5\ \ Facets in the equilibrium crystal }
\esection

Another consequence of the convexity properties,
which will next be discussed (following Miracle-Sole 1995a),
concerns the formation of facets in the equilibrium crystal.
The facets of a crystal have certain particular orientations.
Let ${\bf n}_0$ be the normal corresponding to 
one of these orientations, and 
denote by $\tau(\theta,\phi)$
the function $\tau({\bf n})$ expressed in terms of 
the spherical co-ordinates, 
$0\le\theta\le\pi$, $0\le\phi\le 2\pi$, 
of ${\bf n}$, the vector ${\bf n}_0$
being taken as the polar axis.

\medskip

{\tt Theorem 1.5.} {\it 
Assume that the convexity condition is satisfied.
A facet orthogonal to the direction ${\bf n}_0$ appears
in the Wulff shape if, and only if, the derivative
$\partial\tau(\theta,\phi)/\partial\theta$
is discontinuous at the point $\theta=0$,
for all $\phi$. 
Moreover, the one-sided derivatives
$\partial\tau(\theta,\phi)/\partial\theta$,
at $\theta=0^+$ and $\theta=0^-$,
exist, and determine the shape of the facet.}

\medskip

{\it Proof. }
In terms of the function $f$ defined by (1.18),
the Wulff shape ${\cal W}$ is the set of all
${\bf x}=(x_1,x_2,x_3)\in{\bf R}^3$
such that
$$
x_1 y_1 + x_2 y_2 + x_3 y_3 \le f({\bf y}) 
\eqno(1.31)$$
for every ${\bf y}\in{\bf R}^3$.
If the coordinate axes are placed
in such a way that ${\bf n}_0= (0,0,1)$, 
the plane $x_3=\tau(0)$  
(where $\tau(0)$ is the value of $\tau$ for $\theta=0$)
is a tangent plane to ${\cal W}$.
The facet ${\cal F}$ is the portion of this plane
contained in ${\cal W}$.
These facts follow from the convexity of $f$, 
which implies that $f$
is the support function of ${\cal W}$.
According to (1.31), the facet ${\cal F}$
consists of the points
$(x_1,x_2,\tau (0))\in{\bf R}^3$
such that
$$ 
x_1 y_1 + x_2 y_2 \le f(y_1,y_2,y_3) - y_3 \tau(0) 
= f(y_1,y_2,y_3) - y_3 f(0,0,1) 
\eqno(1.32)$$
for all ${\bf y}=(y_1,y_2,y_3)$.
Or, equivalently, such that
$$ 
x_1 y_1 + x_2 y_2 \le g(y_1,y_2) 
= \inf_{y_3}\ \big(f(y_1,y_2,y_3)-y_3 f(0,0,1)\big) 
\eqno(1.33)$$
Restricting the infimum to $y_3 = 1/\lambda \ge 0$,
and using the positive homogeneity and the
convexity of $f$, one obtains
$$ 
g(y_1,y_2) = \lim_{\lambda \to 0,\lambda \ge 0} (1/\lambda)\  
\big(f(\lambda y_1,\lambda y_2,1) - f(0,0,1)\big) 
\eqno (1.34)$$
This implies that $g$ is a positively homogeneous 
convex function on ${\bf R}^2$.
Define
$$ 
\mu(\phi) = g(\cos\phi,\sin\phi) 
\eqno(1.35)$$
From (1.34), and taking $\lambda =\tan\theta$, one gets
$$\eqalignno{
\mu(\phi) 
&= \lim_{\theta\to0,\theta\ge0} (1/\sin\theta)\  
\big(f(\sin\theta \cos\phi,\sin\theta \sin\phi,\cos\theta)
-\cos\theta f(0,0,1)\big) \cr
&= \lim_{\theta \to 0,\theta \ge 0}(1/\theta)\  
\big(\tau(\theta,\phi) - \tau(0)\big) 
=(\partial/\partial\theta)_{\theta=0^+}\tau(\theta,\phi)
&(1.36)\cr}$$
Similarly
$$
\mu(\phi+\pi) = g(-\cos\phi,-\sin\phi) 
=-(\partial/\partial\theta)_{\theta=0^-}\tau(\theta,\phi)
\eqno(1.37)$$
Equations (1.33) and (1.35) show that both one-sided 
derivatives of exit and,
from the convexity of $g$, it follows that  
$$
\mu(\phi+\pi)\le\mu(\phi) 
\eqno(1.38)$$
Thus, the hypothesis of the discontinuity of the 
derivative $\partial\tau/\partial\theta$, 
at $\theta=0$, implies the strict inequality in (1.38) 
and shows that the convex set ${\cal F}$, given by (1.33),
has a non-empty interior.
This ends the proof of the Theorem.

\medskip

Notice that, according to condition (1.33), the shape of 
the facet is given by  
$$
{\cal F} =\left\{{\bf x}\in{\bf R}^2 :
{\bf x}\cdot{\bf m}\leq\mu({\bf m})\right\}
\eqno(1.39)$$
where the inequality is assumed for every unit vector
${\bf m}=(\cos\phi,\sin\phi)$.

If $\tau(\theta,\phi)=\tau(\pi-\theta,\phi)$ 
the crystal shape is reflection symmetric 
with respect to the plane $x_3=0$. 
Then $\mu({\bf m}) = \mu(-{\bf m})$
(assuming also the symmetry with respect to the origin),
and the facet ${\cal F}$
has a center at the point $P_0 = (0,0,\tau (0))$.
This is not always the case, however, and the point $P_0$ 
can also be outside ${\cal F}$.

\section
{\tt 1.6\ \ Crystal on a wall}
\esection

The variational problem, considered in section 1.1, 
can be solved by purely geometrical means also in
the presence of walls. 
We refer the reader the to paper by Kotecky and Pfister 
(1994) for a general analysis of these situations, 
and consider here, as a basic illustration, the case of
a crystal on a single wall (Winterbottom 1967). 

In this case, 
the surface tension of the boundary of the phase $c$
in contact with the medium differs from that arising
with the wall $w$.
We suppose that the wall is described by a plane
perpendicular to the unit vector ${\bf n}_0$, and that 
$$
E=\left\{{\bf x}\in{\bf R}^d : {\bf x}\cdot{\bf n}_0
<a\right\}
\eqno(1.40)$$
is the half-space where we have the phases $c$ and $m$.
The surface tension between the phases $c$ and $m$
is denoted by $\tau({\bf n})$.
The relevant physical quantity in this problem 
is the difference
$$
\sigma({\bf n}_0)=
\tau_{cw}({\bf n}_0)-\tau_{mw}({\bf n}_0)
\eqno(1.41)$$
where $\tau_{cw}$ and $\tau_{mw}$ are the surface
free energies of the phase $c$ with the wall
and of the phase $c$ with the wall, respectively.
Since $\sigma({\bf n}_0)$ is a difference of free 
energies, it may be positive or negative.

When $\sigma({\bf n}_0)\ge\tau({\bf n}_0)$
we have a drying situation: in equilibrium, it is
preferable that the medium $m$ occupies the place
between the wall and the phase $c$, and consequently
the phase $c$ is not in contact with the wall.
On the other hand, when 
$\sigma({\bf n}_0)\le-\tau({\bf n}_0)$
we have a (complete) wetting situation: in equilibrium, 
the phase $c$ forms a layer between the wall and 
the medium $m$.

In all other cases we speak of partial drying or 
partial wetting.
Then, if $B$ is the shape of the phase $c$, 
the total surface free energy is given by
$$
\tau(\partial B)=
\vert\partial B\cap\partial E\vert\,\sigma({\bf n}_0)
+ \int_{\xi\in\partial B\cap E}\tau({\bf n}(\xi))
ds_\xi
\eqno(1.42)$$
where the first term is the contribution of the boundary 
of the crystal along the wall and the second is the
integral over the remaining part of the boundary of the 
crystal. 
The solution of the variational problem for 
the functional (1.42) is the Winterbottom shape 
${\cal W}$, and can be obtained as follows.
One first constructs the Wulff shape ${\cal W}_\tau$
associated to the function $\tau({\bf n})$
which corresponds to the ideal shape of the free
crystal (formula (1.2)). 
Then we take the intersection of this set with the 
half space defined by the plane orthogonal to
${\bf n}_0$, at a distance $\sigma({\bf n}_0)$ from
the origin, i.e.,  
$$
{\cal W} = {\cal W}_\tau \cap
\left\{{\bf x}\in{\bf R}^d : {\bf x}\cdot{\bf n}_0
<\sigma({\bf n}_0)\right\}
\eqno(1.43)$$ 
We notice that the set ${\cal W}$ is a convex body, 
still defined according to formula (1.2), 
but with the surface tension 
being replaced by $\sigma({\bf n}_0)$ when ${\bf n}_0$ 
is the oriented normal to the interface. 
Thus, equation (1.43) follows, by arguing as in the 
proof of Theorem 1.1, from the isoperimetric inequality. 

\section
{\tt 1.7\ \ Stability of the Wulff solution}
\esection

Let us finally comment on the stability of the Wulff shape 
as the solution to the variational problem of section 1.1.
In the two-dimensional case,  
in addition to the isoperimetric inequality, 
the stability of the minimum in Theorem 1.1 can be 
controlled. 
The following classical inequality, known as the Bonnensen
inequality, gives an estimate from below of the 
isoperimetric defect in the case of the circle.
Suppose that a closed curve of length $L$ bounds a plane
domain $B$, of area $F$, and suppose that $r$ and $R$ 
are the radii of the incircle and the circumcircle, i.e.,
the largest circle contained in $B$ and the 
smallest circle containing $B$.
Then  
$$
L^2-4\pi F\ge \pi^2(R^2-r^2)
\eqno(1.44)$$
In the general case, when ${\cal W}$ is a convex body, 
a similar inequality for the isoperimetric defect 
$$
\tau_{\cal W}(\partial B)^2 - 4\ 
\vert{\cal W}\vert\,\vert B\vert
\eqno(1.45)$$
has been proved by Dobrushin et al.\ (1992). 

There are no estimates of this kind in the three  
dimensional case.
In two dimensions we have that if 
$\tau_{\cal W}(\partial B) - 
\tau_{\cal W}(\partial {\cal W})$ 
tends to zero, 
and $\vert B\vert=\vert{\cal W}\vert$, 
then the Hausdorff distance 
$$
{\dist}_H(\partial{\cal W},B')= 
\max\{\sup_{x\in\partial{\cal W}}\dist(x,\partial B),
\sup_{x\in\partial B}\dist(x,\partial{\cal W})\}
\eqno(1.46)$$ 
between the boundaries of ${\cal W}$ and of 
a suitably chosen translated $B'$ of $B$, tends also
to zero.
This is true in three dimensions if both
${\cal W}$ and $B$ are convex bodies, because of
Diskant inequalities (see for instance Burago and 
Zalgaller 1988), and some appropriate distance could also 
be suggested between general ${\cal W}$ and $B$,  
in order to have the stability of the Wulff solution. 
However, the smallness of the difference
$\tau_{\cal W}(\partial B) - 
\tau_{\cal W}(\partial{\cal W})$ 
does not imply anymore the smallness of the Hausdorff 
distance between $\partial B'$ and $\partial{\cal W}$.  
This is because a ``hair'' attached to the Wulff surface 
does not contribute significantly to the total surface 
free energy. 
From the point of view of thermodynamics, the Wulff shapes 
with long hairs appear to have almost the same free
energy as the equilibrium shape.

\chapter
{\ttt 2\ \  Interfaces in statistical mechanics }
\echapter

In this and the following sections we study some aspects
of the theory of the coexistence of phases in statistical 
mechanics. 

In the first approximation one can model the interatomic 
forces in a crystal by a lattice gas. 
These systems consist of lattice cells which may be either 
empty or occupied by a single particle.
In a typical two-phase equilibrium state there is a 
dense component, which can be identified as the crystal 
phase, and a dilute phase, which can be identified as the 
vapor phase. 
The underlying lattice structure implies that the 
crystal phase is anisotropic, while this assumption, 
though unrealistic for the vapor phase, should be 
immaterial for the description of the crystal-vapor 
interface. 
As an illustrative example of such systems, 
the ferromagnetic Ising model will be considered.

The Ising model is defined on the $d$-dimensional 
cubic lattice ${\cal L}={\bf Z}^d$, 
with configuration space $\Omega = \{-1,1\}^{\cal L}$. 
The value $\sigma(i)$ is the spin at the site $i$.  
The occupation numbers 
$n(i)=(1/2)(\sigma(i)+1$, which take the values $0$ or $1$, 
give the lattice gas version of this model.
The energy of a configuration
$\sigma_{\Lambda} = \{\sigma(i),i\in\Lambda\}$, 
in a finite box $\Lambda\subset{\cal L}$,
under the boundary conditions ${\bar\sigma}\in\Omega$, 
is
$$
H_{\Lambda}(\sigma_{\Lambda}\mid{\bar\sigma})
= - \sum_{\langle i,j\rangle\cap\Lambda\not=\emptyset}
\sigma(i)\sigma(j) 
\eqno(2.1)$$
where 
$\langle i,j \rangle$ are pairs of nearest neighbour sites 
and $\sigma(i) = {\bar\sigma}(i)$ if 
$i\not\in\Lambda$.
The partition function, at the inverse temperature 
$\beta=1/kT$, is given by
$$
Z^{\bar\sigma}(\Lambda)
=\sum_{\sigma_{\Lambda}}\exp \big(-\beta
H_{\Lambda}(\sigma_{\Lambda}\mid{\bar\sigma})\big)
\eqno(2.2)$$
The following limit, which is independent of the 
boundary conditions, 
$$
f(\beta)=\lim_{\Lambda\to\infty}- {1\over{\beta
\vert\Lambda\vert}}\ln\, Z^{\bar\sigma}(\Lambda)
\eqno(2.3)$$
defines the free energy per unit volume. 

It is known that this model presents, at low temperatures
$T<T_c$, where $T_c$ is the critical temperature, 
two distinct thermodynamic pure phases,
a positively and a negatively magnetized phase 
(or a dense and a dilute phase in the lattice gas 
language).
This means two extremal translation invariant Gibbs states, 
which correspond to the limits, when $\Lambda\to\infty$,
of the finite volume Gibbs measures
$$
Z^{\bar{\sigma}}(\Lambda) ^{-1} \exp
\big( - H_{\Lambda}(\sigma_{\Lambda}\mid\bar{\sigma})\big)
\eqno(2.4)$$ 
with boundary conditions ${\bar\sigma}$ respectively 
equal to the ground configurations $(+)$ and $(-)$, 
such that ${\bar\sigma}(i) = 1$ 
and ${\bar\sigma}(i) = -1$, for all $i\in{\cal L}$. 
On the other side, if $T\ge T_c$, 
then the Gibbs state is unique 
(see, for instance, Ruelle 1969, Gallavotti 1972b, 
Miracle-Sole 1976).

Each configuration inside $\Lambda$ can be geometrically 
described by specifying the set Peierls contours, 
which indicate the boundaries between the regions of 
spin $1$ and the regions of spin $-1$.   
Unit square surfaces are placed 
midway between nearest-neighbour pairs of sites $i$ 
and $j$, and perpendicular to these bonds, if 
$\sigma(i)\sigma(j)=-1$. 
The connected components of this set are 
the Peierls contours. 
Under the above boundary conditions $(+)$ and $(-)$, 
they form a set of closed polyhedrical surfaces. 
The contours can be viewed as defects, or excitations,
with respect to the ground states of the system 
(the constant configurations $1$ and $-1$), 
and are a basic tool for the investigation
of the model at low temperatures.

\section
{\tt  2.1\ \ The surface tension}
\esection

In order to study the interface between the two pure phases
one needs to construct a state describing the coexistence
of these phases.
Let $\Lambda$ be a parallelepiped of sides 
$L_1,\dots,L_d$, parallel to the axes,
and centered at the origin of ${\cal L}$, 
and let ${\bf n}=(n_1,\dots,n_d)$ 
be a unit vector in ${\bf R}^d$, such that $n_d\ne 0$. 
Introduce the mixed boundary conditions $(\pm,{\bf n})$, 
for which ${\bar\sigma}(i)=1$ if $i\cdot{\bf n}\geq 0$, 
and ${\bar\sigma}(i)=-1$ if $i\cdot{\bf n}<0$. 
These boundary conditions force the system to produce a
defect going transversely through the box $\Lambda$,
a big Peierls contour that can be interpreted as a 
microscopic interface.
The other defects that appear above and below the 
interface can be described by closed contours
inside the pure phases.

The free energy, per unit area, due to the presence of the 
interface, is the surface tension. 
It can be defined by
$$
\tau({\bf n})=
\lim_{L_1,\dots,L_{d-1}\to\infty}\, \lim_{L_d
\to\infty} \, -{{n_d}\over{\beta L_1\dots L_{d-1}}}
\ln\, {Z^{(\pm,{\bf n})}(\Lambda)\over 
Z^{(+)}(\Lambda)} 
\eqno(2.5)$$ 
Notice that in this expression the volume contributions 
proportional to the free energy of the coexisting phases, 
as well as the boundary effects, cancel, and only 
the contributions to the free energy of the interface 
are left. 

\medskip

{\tt Theorem 2.1 } {\it 
The thermodynamic limit $\tau ({\bf n})$, 
of the interfacial free energy per unit area, exists, 
and is a non negative bounded function of ${\bf n}$. 
Its extension by positive homogeneity,  
$f({\bf x})=\vert{\bf x}\vert\, \tau({\bf x}/ \vert 
{\bf x}\vert)$, 
is a convex function on ${\bf R}^d$.}

\medskip

A proof of these statements has been given by 
Messager et al.\ (1992) using correlation inequalities 
(this being the reason for their validity for all $\beta$). 
In fact the validity of this Theorem has been proved 
for a large class of lattice systems.

Moreover, for the Ising model we know, 
from Bricmont et al.\ (1980), Lebowitz and Pfister 
(1981) and the convexity condition, 
that $\tau ({\bf n})$ is strictly positive for
$T<T_c$ and that it vanishes if $T\ge T_c$.


\section
{\tt  2.2\ \ Interfaces at low temperatures }
\esection

In Appendix A we give a r\'esum\'e of the theory 
of cluster expansions. 
We are going to apply this theory to study the low 
temperature properties of the system. 
Let us first describe the cluster theory of a pure phase. 
We have seen above that a configuration in the 
box $\Lambda$, with $(+)$ boundary conditions, is 
characterized as a set $X=\{\gamma_1,\dots,\gamma_n\}$ 
of closed mutually disjoint contours. 
It is easy to see that, up to a constant, 
the energy of the configuration is equal to twice the 
total area of the contours. 
Besides the mutual exclusion there is no other 
interaction between the contours. 
If to each contour we attribute the weight
$\phi(\gamma)=\exp(-2\beta\vert\gamma\vert)$, where  
$\vert\gamma\vert$ is the area of the contour $\gamma$,  
we get, for the partition function $Z^{(+)}(\Lambda)$,  
an expression like equation (A.1), in Appendix A.  
The system can be viewed in terms of contours as 
a system of polymers, with weight $\phi(\gamma)$,
and the convention that two contours are compatible 
if they do not intersect. 

Let us next consider the convergence condition,  
Theorem A.2 in Appendix A. 
We introduce 
$\mu(\gamma)=\exp(-b\vert\gamma\vert)$ 
and remark that the number of contours cutting 
a given lattice bond and of area $\vert\gamma\vert=n$ 
is bounded by $K^n$, where $K$ is certain constant 
(one can take $K=3$ using appropriate definitions). 
Then, we have 
$$
\sum_{\gamma'\not\sim\gamma}\mu(\gamma') \le
\vert\gamma\vert\sum_{n}K^n e^{-bn} \le 
\vert\gamma\vert (1-K e^{-b})^{-1}
\eqno(2.6)$$ 
This shows that the convergence condition (A.6) 
is satisfied if the temperature is sufficiently low, 
i.e., there exist a constant $\beta_0$ such that (A.6) 
is satisfied if $\beta\ge\beta_0$. 
Using formula (A.4) and estimate (A.7), one can then 
derive a convergent series expansion for the free energy
of the system. 
Also, from (A.4) and (A.7), it follows that 
$$
Z^{(\pm,{\bf n})}(\Lambda) / Z^{(+)}(\Lambda)
= \sum_{{\cal I}} e^{-2\beta\vert{\cal I}\vert}
\exp\big(-\sum_{X\cap{\cal I}\ne\emptyset} 
a^{\rm T}(X)\phi^X\big) 
\eqno(2.7)$$
where the first sum runs over all microscopic interfaces 
${\cal I}$, compatible with the boundary 
conditions. 
The sum in the exponential runs over all clusters
of contours that intersect ${\cal I}$. 
Each term in the first sum gives a weight proportional to 
the probability of the microscopic interface. 
From this, a representation of the microscopic interface 
itself as a polymer system, can be derived. 

Consider now the three-dimensional Ising model with   
the $(\pm,{\bf n}_0)$ boundary conditions,  
associated to the direction ${\bf n}_0 = (0,0,1)$, 
which define inside the box $\Lambda$ a horizontal
interface. 
At low temperatures $T>0$, we expect the microscopic 
interface corresponding to these boundary conditions,
which at $T=0$ coincides with the plane $i_3=-1/2$,
to be modified by small deformations.
The microscopic interface can then be described  
by means of its defects, or excitations, 
with respect to the interface at $T=0$.
These defects, called walls, form the boundaries
(which may have some width), between the smooth plane 
portions of the interface.
In this way the interface structure, with its probability 
distribution defined by equation (2.7), 
may be interpreted as a ``gas of walls'' on 
a two-dimensional lattice. 
There is an interaction between the walls,  
coming from the fact that the interface is surrounded
by the $(+)$ phase from above and the $(-)$ phase
from below. 
But the exponential function in equation (2.7),
gives a mathematical description of this interaction 
in terms of clusters of contours. 
This allows us to define a ``gas of aggregates of walls 
and clusters'' as a polymers system on the two-dimensional 
lattice
(see the original articles quoted below and, for instance,
Miracle-Sole 1995a, Appendix, for a short review).

Dobrushin (1972) proved the dilute character of this 
gas at low temperatures, which means that the interface 
is essentially flat (or rigid).
The considered boundary conditions yield indeed a
non translation invariant Gibbs state.
Furthermore, cluster expansion techniques have been 
applied by Bricmont et al.\ (1979),
to study  the interface structure in this case
(see also Holicky et al.\ 1988).

The same analysis applied to the two-dimensional model
shows a different behavior at low temperatures.
In this case the walls belong to a one-dimensional lattice,
and Gallavotti (1972) proved that the microscopic 
interface undergoes large fluctuations of order 
$\sqrt{L_1}$.
The interface does not survive in the thermodynamic
limit, $\Lambda\to\infty$,
and the corresponding Gibbs state is translation invariant.
Moreover, the interface structure can be studied by means 
of a cluster expansion for any orientation of the interface
(see also Bricmont et al.\ 1981).

In the three-dimensional case, the description of the 
microscopic interface for any orientation ${\bf n}$, 
leads to very difficult problems of random surfaces.    
It is possible, however, by introducing a new polymer 
system, to analyze also the interfaces which are near 
to the particular orientations ${\bf n}_0$,  
discussed above.
This problem, related to the formation of facets 
in the equilibrium crystal, will be treated in section 4.

\vfill\eject

\chapter
{\ttt 3\ \ On the microscopic proof of the Wulff 
construction }
\echapter

In this section we consider a simplified model
of the two dimensional interface. 
We represent the interface as a 
single-valued function over a reference line. 
At each site $i$ of the lattice ${\bf Z}$
an integer variable $h_i$ is assigned 
which indicates the 
height of the interface at this site. 
One obtains a statistical mechanical model by 
assigning an energy to each configuration 
${\bf h} = \{ h_0,h_1,...,h_N \}$,
in the box $0\le i\le N$, of length $N$.
The standard example is
$$ 
H_N ({\bf h}) = \sum_{i=1}^N (1+|h_i-h_{i-1}|) 
\eqno{(3.1)}$$
which corresponds to the solid-on-solid (SOS) model.
Then the energy of a configuration 
coincides with the length of the interface.
The weight of a given configuration,
at the inverse temperature $\beta $,
is proportional to the Boltzmann factor \  
$\exp \big( -\beta H_\Lambda(\phi)\big)$.
Provided that the energy remains
unchanged under the global shift
$h_i$ to $h_i+b$, for all $i$ and all
$b\in{\bf Z}$, there is considerable freedom of 
how to choose the energy function.  
The results below apply, also, to
more general Hamiltonians of this form.

The model provides an approximate
description of an interface separating
two phases at equilibrium,
such as the positively and negatively magnetized
phases of the three-dimensional Ising model.
Actually the SOS model may be obtained as the limit
of the anisotropic Ising model,
when we let the coupling constant,
in the vertical direction, tend to infinity.

We are going to study the statistical mechanics of the 
SOS model, with the global constraint of having a 
specified area between the interface and the horizontal 
axis, and to show the existence of the thermodynamic 
limits and the equivalence of the Gibbs ensembles 
associated with this problem. 
As a corollary, it can be seen that the configurations of 
a large system follow a well defined mean profile with 
probability one, 
the shape of the phase boundary at equilibrium.
This gives a simple alternative microscopic proof of
the validity of the Wulff construction 
for such models,
first established by DeConink, et al.\ (1989).
The fluctuations of the interface around the mean profile
have recently been studied by Dobrushin and Hryniv (1996)
following the same approach. 

\section
{\tt 3.1\ \ Surface tension and Wulff shape}
\esection

We introduce first the Gibbs ensemble which consists 
of all configurations, in the box of length $N$, 
with specified boundary conditions
$h_0=0$ and $h_N=Y$.
The associated partition function is given by
$$ 
Z_1(N,Y) = \sum_{\bf h} e^{-\beta H({\bf h})}
\ \delta(h_0)\delta(h_N-Y) 
\eqno{(3.2)}$$
where the sum runs over all configurations in the box
and $\delta(t)$ is the discrete Dirac delta
($\delta (t)=1$ if $t=0$ and $\delta(t)=0$ otherwise).
We define the corresponding free energy per site as the limit
$$ 
\tau_p(y) = \lim_{N\to\infty}-{1\over{\beta N}} \ln Z_1(N,yN) 
\eqno{(3.3)} $$
where $y=-\tan\theta$, the slope of the interface,
is a real number.
This free energy is called the projected surface tension.
The surface tension,
which represents the interfacial free energy per unit length
of the mean interface, is
$$ 
\tau (\theta ) = \cos \theta \  \tau _p (- \tan \theta ) 
\eqno{(3.4)}$$

We introduce a second Gibbs ensemble,
conjugate to the previous ensemble,
whose partition function, in the box of length $N$,
is given by
$$ 
Z_2 (N,x) = \sum_{\bf h} e^{-\beta H({\bf h})}
e^{\beta x h_N}\ \delta (h_0) 
\eqno{(3.5)}$$
where $x\in{\bf R}$, 
replaces as a thermodynamic parameter the slope $y$. 
We define the associated free energy as
$$ 
\varphi(x)=\lim_{N\to\infty}-{1\over{\beta N}}\ln Z_2(N,x) 
\eqno{(3.6)}$$

\medskip

{\tt Theorem 3.1.} {\it
Limits (3.3) and (3.6), which define the above free 
energies, exist.
The first, $\tau_p$, 
is a  convex even function of $y$.
The second, $\varphi $,
is a concave even function of $x$.
Moreover, $\tau_p$ and $-\varphi $ are conjugate
convex functions, i.e.,
they are related by the Legendre transformationsÊ}
$$\eqalignno{
-\varphi(x) &=\sup_y\  \big(x y-\tau_p(y)\big) \cr
\tau_p(y) &=\sup_x\  \big( x y + \varphi(x)\big) 
&{(3.7)}\cr}$$

\medskip

The validity of the above statements follows from the
subadditivlty property 
$$
\ln Z_1(N_1 + N_2,y(N_1 + N_2)) \ge 
\ln Z_1(N_1,yN_1)+\ln Z_1(N_2,yN_2)
\eqno(3.8)$$ 
which is straightforward, and standard arguments 
in the theory of the thermodynamic limit 
(Ruelle 1969, Galgani et al.\ 1971).
See also, for instance, Messager et al.\ (1992)
for a proof of these statements in a more general setting.

The convexity of $\tau_p$ is equivalent to the fact that
the surface tension $\tau$ satisfies 
the pyramidal inequality (see section 1.3).
Relations (3.7) between the free energies express
the thermodynamic equivalence of the two ensembles 
(3.2) and (3.5). 
These relations imply that the curve
$z = \varphi (x)$ gives,
according to the Wulff construction,  
or the equivalent Andreev construction,
the equilibrium shape of the crystal
associated to our system (see section 1.4).

The function $\varphi (x)$ defined by (3.6) is easily 
computed by summing a geometrical series. 
One introduces the difference variables
$$ 
n_i = h_{i-1} - h_i 
\eqno{(3.9)}$$
for $i = 1,...,N $, so that
the partition function factorizes and one obtains
$$ 
\varphi (x) = 1
- \beta ^{-1} \ln \sum _{n}
e^{ - \beta |n| + \beta x n} 
\eqno{(3.10)}$$
The explicit form of this function is
$$ 
\varphi (x) = 1-\beta^{-1}
\ln{{\sinh\beta}\over{\cosh\beta - \cosh\beta x}} 
\eqno{(3.11)}$$ 
if $-1< x <1$, and $\varphi(x) = -\infty$ otherwise.

\section
{\tt 3.2\ \  The volume constraint}
\esection

We next define two new Gibbs ensembles. 
In the first of these ensembles we consider the 
configurations such that $h_N = 0$, which have a 
specified height at the origin $h_0 = M$
and which have a specified volume (area) $V$
between the interface and the horizontal axis,
this volume being counted negatively for negative heights, 
$$ 
V = V({\bf h}) = \sum _{i=0}^N h_i 
\eqno{(3.12)}$$
The corresponding partition function is
$$ 
Z_3 (N,V,M) = \sum _{\bf h} e^{- \beta H({\bf h})}
\ \delta (h_N) \delta (V({\bf h}) - V)
\delta (h_0 - M) 
\eqno{(3.13)} 
$$
where $V$ and $M$ are understood as their integer parts 
when they do not belong to ${\bf Z}$.
The second ensemble is the conjugate ensemble of (3.13). 
Its partition function is given by
$$ 
Z_4 (N,u,\mu) = \sum_{\bf h} e^{-\beta H({\bf h})}
\ e^{\beta u (V({\bf h})/N) + \beta \mu h_0 }
\ \delta (h_N) 
\eqno{(3.14)}$$
where $u\in{\bf R}$ and $\mu\in{\bf R}$
are the conjugate variables.
Our next step will be to prove the existence of the
thermodynamic limit for these ensembles and their
equivalence in this limit.

\medskip

{\tt Theorem 3.2. } {\it 
The following limits exist 
$$\eqalignno{
\psi_3(v,m) 
&=\lim_{N\to\infty} -{1\over{\beta N}}\  \ln Z_3(N,v N^2,mN) 
&{(3.15)} \cr
\psi_4(u,\mu) 
&= \lim_{N\to\infty}-{1\over{\beta N}}\  \ln Z_4(N,u,\mu) 
&{(3.16)} \cr}
$$
and define the free energies per site associated to 
the considered ensembles. 
Moreover, $\psi_3$ and $-\psi_4$ are conjugate 
convex functions }
$$\eqalignno{
-\psi_4 (u,\mu) &=
\sup_{v,m}\  \big( u v + \mu m - \psi_3(v,m)\big) \cr
\psi_3 (v, m) &=
\sup_{u,\mu}\  \big( u v + \mu m + \psi_4 (u,\mu)\big) 
&{(3.17)}\cr}$$

\medskip 

The crucial observation for proving this Theorem
is the subadditivity property given below. 
Then we adapt known arguments (Ruelle 1969, 
Galgani et al.\ 1971)
in the theory of the thermodynamic limit. 
For a more detailed proof see Miracle-Sole and 
Ruiz (1994) or Dobrushin and Hryniv (1996).

\medskip

{\tt Proposition. }  
The partition function $Z_3$
satisfies the subadditivity property 
$$\eqalignno{
Z_3 (2N,2&(V'+ V''),M'+ M'') \ge \cr
&Z_3 (N,V',M')\  Z_3 (N,V'',M'') \  
e^{- 2 \beta |M''| / (2N-1) } 
&{(3.18)}\cr}$$

\medskip

{\it Proof. }
In order to prove this property we
associate a configuration ${\bf h}$
of the first system in the box of length $2N$,
to a pair of configurations ${\bf h}'$ and ${\bf h}''$
of the system in a box of length $N$, as follows
$$
\eqalignno{
h_{2i} &=
h'_{i} + h''_{i} \; , i=0,\dots,N \cr
h_{2i-1} &=
h'_{i-1 } + h''_{i} \; , i=1,\dots, N 
&{(3.19)}\cr}
$$
Then 
$h_{2N} = h'_{N} + h''_{N} = 0$,  
$h_{0} = h'_{0} + h''_{0} = M' + M''$ 
and
$$\eqalignno{
V({\bf h}) &=
2 \sum_{i=1}^N h'_i + \sum_{i=0}^N h''_i 
+ \sum_{i=1}^N h''_i \cr
&= 2\  \big(V({\bf h'}) + V({\bf h''})\big) - M'' 
&(3.20)\cr}$$
This shows that the configuration $\bf h$ belongs to
the partition function in the right hand side of (3.18).
Since 
$$ 
H_{2N}({\bf h}) = H_N ({\bf h'}) + H_N ({\bf h''})
\eqno(3.21)$$
because $n_{2i}=n'_i$ and $n_{2i-1} =n''_i$, 
as follows from (3.19), we get 
$$
Z_3 (N,V',M')\  Z_3 (N,V'',M'') \le 
Z_3 (2N, 2(V' + V'') - M'', M' + M'')
\eqno(3.22)$$
Then we use the change of variables
$$\eqalignno{
&\tilde{h}_i = h_i + (M''/(2N-1)),\ i=1,\dots,2N-1,\cr 
&\tilde{h}_0 = h_0,\  \tilde{h}_{2N} = h_{2N}=0
&(3.23)\cr}$$ 
which gives
$$
Z_3 ( 2N, V - M'', M )\le e^{ 2 \beta |M''| / (2N-1)}
\ Z_3 ( 2N, V, M)
\eqno(3.24)$$
to conclude the proof.   

\section
{\tt 3.3\ \ The Wulff construction}
\esection

The free energies $\psi_3$ and $\psi_4$, associated to the 
system with the volume constraint, 
can be expressed in terms of the functions
$\varphi$ and $\tau_p$ introduced in section 3.2.
We consider first the function $\psi_4$.
In terms of the difference variables (3.9), 
we have 
$$
V({\bf h}) = \sum_{i=0}^N h_i = \sum _{i=1}^N \; i n_i 
\eqno(3.25)$$
and, therefore,  
$$ 
Z_4 (N,u,\mu) 
= \prod _{i=1} ^N \Big( \sum_{n_i}
e^{-\beta |n_i| + \beta (u/N) i n_i + \beta\mu n_i} \Big) 
\eqno(3.26)$$
Taking expression (3.10) into account it follows
$$
Z_4 (N,u,\mu) 
= \exp\Big(-\beta\sum_{i=1}^N
\varphi \big({u\over N} i + \mu \big)\Big) 
\eqno(3.27)$$
and
$$
\psi_4 (u,\mu) = 
\lim_{N \to \infty} {1 \over N} 
\sum_{i=1}^N \varphi\big({ u \over N} i + \mu \big) 
= \lim_{N \to \infty} {1 \over u} 
\sum_{i=1}^N {u \over N} 
\varphi\big({ u \over N} i + \mu \big) 
\eqno(3.28)$$
which, writing the Riemann sum as an integral, implies  
$$
\psi_4(u,\mu) =
{1\over u} \int_0^u \varphi(x + \mu) dx 
\eqno(3.29)$$

Next we consider the free energy $\psi _3$. 
This function is determined by the Legendre 
transform (3.17).
The supremum over $u,\mu$ is obtained for the values  
$u_0,\mu_0$, 
for which the partial derivatives of the right hand side 
of equation (3.17) are zero:
$$\eqalignno{
v + (\partial\psi_4/ \partial u) (u_0,\mu_0) &= 0 \cr 
m + (\partial\psi_4/ \partial \mu) (u_0,\mu_0) &= 0 
&(3.30)\cr}$$ 
That is, for $u_0,\mu_0$  which satisfy 
$$\eqalignno{
{1\over{u_0^2}} 
\int_0^{u_0} \varphi(x + \mu_0) dx 
- {1\over{u_0}} \varphi(\mu_0 + u_0) &= v
&(3.31) \cr
{1\over u_0} 
\big(\varphi(\mu_0 ) - \varphi(\mu_0 + u_0)\big) &= m
&(3.32)\cr}
$$
Then, from (3.17), (3.28), (3.31) and (3.32),
we get
$$
\psi_3 (v,m) = 2 \psi_4 (u_0,\mu_0) -
{{1}\over{u_0}} \big((\mu_0 + u_0)\varphi(\mu_0 + u_0)
- \mu_0 \varphi(\mu_0)\big)
\eqno(3.33)$$
But, using relation (3.7) under the form
$$
\varphi(x) = 
x \varphi'(x) + \tau_p (\varphi'(x)) 
\eqno(3.34)$$
in (3.29) and integrating by parts, we get 
$$
2 \psi_4 (u_0,\mu_0) = {{1}\over{u_0}} 
\int_{\mu_0}^{\mu_0 + u_0} \tau_p (\varphi'(x)) dx 
+ {{1}\over{u_0}} \big((\mu_0 + u_0) \varphi(\mu_0 + u_0)
- \mu_0 \varphi(\mu_0)\big)
\eqno(3.35)$$
Finally, this equation together with (3.33), implies 
$$
\psi_3(v,m) =
{{1}\over{u_0}}    
\int_{\mu_0}^{\mu_0 + u_0}\tau_p(\varphi'(x)) dx 
\eqno(3.36)$$

To interpret these relations, let us consider the
graph of the function $z=\phi(x)$ and an arc $BC$
of this curve between the points $B$ and $C$ 
having the abcises $\mu_0$ and $\mu_0+u_0$, respectively.
We draw the vertical line $x=\mu_0$ passing through 
the point $B$, and the horizontal line 
$z=\phi(\mu_0+u_0)$ passing through the point $C$.
Let $A$ be the point where these two lines intersect.
We observe that the right 
hand side of (3.31) represents the area $ABC$ 
divided by $\overline{AC}^2$.
Therefore, the values $u_0$, $\mu_0$, which solve   
(3.31) and (3.32), 
are obtained when this area is equal to $v$, with the 
condition coming from (3.32), that the slope 
$\overline{AB}/\overline{AC}$, is equal to $m$.
Then, according to (3.36), the free energy $\psi_3(v,m)$
is equal to the integral of the surface tension along 
the arc $BC$, of the curve $z=\varphi(x)$, 
divided by the same scaling factor $\overline{AC}=u_0$.

We conclude that, for large $N$, the configurations 
of the SOS model, with a prescribed area $vN^2$,
follow a well defined mean profile, that is, 
the macroscopic profile given by the Wulff construction,
with very small fluctuations.
This follows from the fact that the probability of the 
configurations which deviate macroscopically from the mean 
profile is zero in the thermodynamic limit. 
The free energy associated to the configurations which 
satisfy the conditions above, and moreover, 
are constrained to pass through a given point not 
belonging to the mean profile, 
can be computed with the help of equation (3.13).
Indeed, if the coordinates of the point are $(x_1,m_1)$,
then the partition function associated to such set of
configurations is given by the product
$$
Z_3\big(x_1N,V',(m-m_1)N\big)\ 
Z_3\big((1-x_1)N,V'',(m-m_1)N\big)
\eqno(3.37)$$
with moreover the condition $V'+V''=vN^2$.
As a consequence of equations (3.15) and (3.36), and knowing 
from the macroscopic variational problem that
the arc $BC$ is the shape which minimizes the
integral of $\tau_p$, 
we see that the probability of this set of configurations 
tends exponentially to zero as $N\to\infty$.

\section
{\tt 3.4\ \ Remark}
\esection

The mathematically rigorous justification of the 
Wulff construction, 
in the case of the two-dimensional Ising model 
at low temperatures, is due to Dobrushin et al.\ 
(1992, see also Dobrushin et al.\ 1993, for the 
statement of the main result and some ideas of the proof,
and Pfister 1991, for another version of the same proof). 
Their results show that, in the canonical ensemble, 
where the total number of particles 
(or the total magnetization in the language of 
spin systems) is fixed, a (unique) droplet of 
the dense phase, immersed in the dilute phase, 
is formed. 
Its shape, when properly rescaled, obeys the Wulff 
principle. 
Ioffe (1994, 1995) extended these results,  
in the case of the two-dimensional Ising model, 
to all temperatures below the critical temperature.

\chapter
{\ttt 4\ \  Roughening transition and equilibrium shapes}
\echapter

Let us come back to the three-dimensional Ising model. 
We already know, from section 2, that the interface 
orthogonal to a lattice axis is rigid at low temperatures, 
and that the corresponding boundary conditions give rise 
to a non translation invariant Gibbs state. 
It is believed, that at higher temperatures,
but before reaching the critical temperature $T_c$, 
the fluctuations of this interface 
become unbounded when the volume tends to infinity,
so that the corresponding Gibbs state in the
thermodynamic limit is translation invariant.
The interface undergoes a roughening phase
transition at a temperature $T=T_R<T_c$.

Recalling that the rigid interface may be viewed 
as a two-dimensional polymer system (the system of
walls), one might expect that the critical temperature 
$T_c^{d=2}$, of the two-dimensional Ising model, is
relevant for the roughening transition, and that
$T_R$ is somewhere near $T_c^{d=2}$. 
Indeed, approximate methods, used by Weeks et al.\ (1973), 
suggest $T_R\sim 0.53\ T_c$,
a temperature slightly higher then $T_c^{d=2}$. 
Moreover, van Beijeren (1975) proved, 
using correlation inequalities,
that $T_R\ge T_c^{d=2}$.

Since then, however, it appears to be no proof of the 
fact that $T_R < T_c$, i.\ e., that the roughening transition 
for the three-dimensional Ising model really occurs. 

At present one is able to study rigorously the roughening 
transition only for some simplified models of the 
microscopic interface.  
Thus, Fr\"ohlich and Spencer (1981) have proved 
this transition for the SOS (solid-on-solid) model. 
Moreover, several restricted SOS models, 
in which the differences of heights at two 
nearest neighbor sites are restrict to take a few 
number of values, are exactly solvable. 
They present also a roughening transition 
(this theory has been reviewed by Abraham 1986 and 
van Beijeren and Nolden 1987).

From a macroscopic point of view,
the roughness of an interface should be apparent
when considering the shape of the equilibrium crystal 
associated with the system.
One knows that a typical equilibrium crystal at low 
temperatures has smooth plane facets 
linked by rounded edges and corners.
The area of a particular facet decreases as the 
temperature is raised and the facet finally disappears 
at a temperature characteristic of its orientation.
The reader will find information and references
on equilibrium crystals in the review articles by
Abraham (1986), van Beijeren and Nolden (1987), 
Kotecky (1989) and Rottman and Wortis (1984). 

It can be argued  
that the roughening transition corresponds to the 
disappearance of the facet whose orientation is the same 
as that of the considered interface.
The exactly solvable SOS models mentioned above,
for which the function $\tau({\bf n})$ 
has been computed,
are interesting examples of this behavior
(this subject has been reviewed by Abraham 1986, 
Chapter VII, see also Kotecky and Miracle-Sole
1986, 1987a, 1987b). 
This point will be briefly discussed in the next two 
sections, where several results on this subject,   
concerning the three-dimensional Ising model,  
will be reported. 
See Miracle-Sole (1995a) for the proofs and 
a more detailed discussion.

\section
{\tt 4.1\ \  The step free energy}
\esection

The step free energy plays an important role in the problem 
under consideration. 
It is defined, using appropriate boundary conditions,
as the free energy associated with the introduction
of a step of height 1 on the interface.
This quantity can be regarded as an order parameter
for the roughening transition, analogous, in some sense,
to the surface tension in the case of the usual 
phase transitions.
Indeed, Bricmont et al.\ (1982) proved 
that $\tau^{\rm step} > 0$ if $T < T_c^{d=2}$, 
a result analogous to the above mentioned result 
by van Beijeren (1975). 
On the other hand $\tau^{\rm step} = 0$ if $T \ge T_c$
(see, again, Bricmont et al.\ 1982). 

In order to define the step free energy we consider 
a parallelepipedic box $\Lambda$, 
defined as in section 2 above, and introduce the 
$({\rm step},{\bf m})$
boundary conditions, associated to the unit vectors
${\bf m} = (\cos\phi,\sin\phi)\in{\bf R}^2$,
by
$$
{\bar\sigma}(i) = \cases{
1  &if $i>0$  or if $i_3=0$ and $i_1m_1+i_2m_2\ge 0$ \cr
-1 &otherwise \cr}  
\eqno(4.1)$$
Then, the step free energy, for a step
orthogonal to ${\bf m}$ (such that $m_2\ne0$),  
is 
$$ 
\tau^{\rm step}(\phi) =
\lim_{L_1\to\infty}\lim_{L_2\to\infty}\lim_{L_3\to\infty}
- {{\cos \phi}\over{\beta L_1}}\  
\ln\  {{Z^{({\rm step},{\bf m})}(\Lambda )}\over 
{Z^{(\pm,{\bf n}_0)}(\Lambda )}} 
\eqno(4.2)$$
Clearly, this expression represents the residual free
energy due to the considered step, per unit length.

When considering the configurations under the 
$({\rm step},{\bf m})$
boundary conditions, the step may be viewed as a defect
on the rigid interface described in section 2. 
It is, in fact,
a long wall going from one side to the other side 
of the box $\Lambda$.
A more careful description of it can be obtained as follows.
At $T=0$, the step parallel to the axis 
(i. e., for ${\bf m}=(0,1)$)
is a perfectly straight step of height 1.
At a low temperature $T>0$, some deformations appear,
connected by straight portions of height 1.
The step structure, with its probability distribution 
in the corresponding Gibbs state,
can then be described as a ``gas'' of these defects
(to be called step-jumps).  
This system can be represented as a polymer system 
on a one-dimensional lattice.
This description, somehow similar to the description of the
interface of the two-dimensional Ising model used by 
Gallavotti (1972),
is valid, in fact, for any orientation ${\bf m}$ of the step.
It can be shown that the system of step-jumps,  
at low temperatures, satisfies the properties 
which are required for applying the cluster expansion 
techniques.  
In this way, the step structure can be studied. 
Actually, the step-jumps are not independent since
the rest of the system produces an effective interaction
between them.
Nevertheless, this interaction can be treated by means 
of the low temperature expansion, in terms of walls,  
for the rigid interface, applied to the regions of the 
interface lying at both sides of the step.
From this analysis (Miracle-Sole 1995a) 
one gets the following result.

\medskip

{\tt Theorem 4.1. } {\it 
If the temperature is low enough 
(i.e., if $T\le T_0$, where $T_0 > 0$
is a given constant), 
then the step free energy $\tau^{\rm step}({\bf m})$,
exists in the thermodynamic limit, and 
extends by positive homogeneity to a strictly convex function.
Moreover, $\tau^{\rm step}({\bf m})$ 
can be expressed in terms of an analytic function of $T$, 
which can be obtained by means of a 
convergent cluster expansion.}

\medskip

In fact, 
$$ 
\eqalignno{ 
\tau^{\rm step}({\bf m}) =\ \ 
&2 (|m_1|+|m_2|) 
- (1/\beta) \big( (|m_1|+|m_2|) \ln (|m_1|+|m_2|) \cr  
&- |m_1|\ln|m_1| - |m_2|\ln|m_2| \big)
- (1/\beta) \varphi_{\bf m} (\beta)  
&(4.3)\cr}$$
where $ \varphi_{\bf m} $
is an analytic function
of $z=e^{-2\beta}$,
for $|z| \le e^{-2\beta_0}$. 
The first two terms in this expression, 
which represent the main contributions 
for $T\to 0$, 
come from the ground state of the system
under the considered boundary conditions.
The first term can be recognized as the residual energy
of the step at zero temperature
and, the second term, as $-(1/\beta)$ times 
the entropy of this ground state.
The same two terms occur in the surface tension of the
two-dimensional Ising model (see Avron et al.\ 1982
and Dobrushin et al.\ 1992). 
By considering the lowest energy excitations, 
it can be seen that  $\varphi_{\bf m}$
is $O(e^{-4\beta})$, 
and also, that the first term in which this series
differs from the series associated to 
the surface tension of the two-dimensional Ising model, 
is $O(e^{-12\beta})$. 

\section
{\tt 4.2\ \  The shape of a facet }
\esection

We have seen, in section 1.5, that the appearance of a facet 
in the equilibrium crystal shape is related, 
according to the Wulff construction, to the existence 
of a discontinuity in the derivative of the surface
tension with respect to the orientation. 
In the case of the three-dimensional Ising model, 
a first result concerning this point,   
was obtained by Bricmont et al.\ (1986).  
These authors proved
a correlation inequality which establish
$\tau^{\rm step}$ 
as a lower bound to the one-sided derivative
$\partial\tau(\theta) / \partial\theta$ at $\theta=0^+$ 
(here $\tau^{\rm step}=\tau^{\rm step}(0,1)$ 
and $\tau(\theta)=\tau(0,\sin\theta,\cos\theta)$).    
Thus $\tau^{\rm step}>0$ implies a kink in $\tau(\theta)$
at $\theta=0$ and, therefore, a facet is expected. 

In fact, $\tau^{\rm step}$ should be equal to this 
one-sided derivative. 
This is reasonable, since the increment in surface tension 
of an interface tilted by an angle $\theta$, 
with respect to the surface tension of the rigid interface,
can be approximately identified, for $\theta$ small, 
with the free energy of a collection of steps.   
The density of the steps being proportional to $\theta$,   
the distance between them becomes very large 
when $\theta$ is very  small. 
But, if the interaction between the steps can be neglected, 
the free energy of the whole collection of steps  
can be approximated by the sum of the individual 
free energies of the steps.

With the help of the methods described in section 4.1,
it is possible to study these free energies and to control 
the approximations involved in the problem. 
The following result can then be derived
(see Miracle-Sole 1995a for the proof).

\medskip

{\tt Theorem 4.2. } {\it
For $T<T_0$, we have
$$
\partial\tau(\theta,\phi)/\partial\theta |_{\theta=0^+}
= \tau^{\rm step}(\phi)   
\eqno(4.4)$$
i.\ e., the step free energy equals the one-sided angular
derivative of the surface tension.} 

\medskip

It is natural to expect that this equality is true for
any $T$ less than $T_R$, and that for $T\ge T_R$,  
both sides in the equality vanish,
and thus, the disappearance of the facet is involved.
However, the condition that the temperature 
is low enough is important here.
Only when it is fulfilled we have the full control
on the equilibrium probabilities that is needed in the proofs. 

The above relation, together with Theorem 1.5, 
implies that the shape of the facet is given 
by the two-dimensional Wulff construction
applied to the step free energy $\tau^{\rm step}({\bf m})$.
Namely,
$$
{\cal F} =\left\{ {\bf x}\in{\bf R}^2 :
{\bf x}\cdot{\bf m}\leq\tau^{\rm step}({\bf m}) \right\}
\eqno(4.5)$$ 
where the inequality is assumed for every unit vector 
${\bf m}$ in ${\bf R}^2$. 
Then, from the properties of $\tau^{\rm step}$ mentioned 
above, it follows that
the facet has a smooth boundary without straight segments and,
therefore, that the crystal shape presents rounded edges 
and corners.

\chapter
{\ttt Appendix A : Cluster expansions}
\echapter

Let ${\cal P}$ be a countable or a finite set, 
the elements of which will be called polymers. 
Let ${\cal I}\subset{\cal P}\times{\cal P}$ be a reflexive and 
symmetric relation.
We say that the elements $\gamma_1,\gamma_2\in{\cal P}$ are
incompatible if $(\gamma_1,\gamma_2)\in{\cal I}$, 
and we will also write $\gamma_1\not\sim\gamma_2$.
If $(\gamma_1,\gamma_2)\not\in{\cal I}$ we say that the two
polymers are compatible. 

For each finite set $X\subset{\cal P}$, 
we define $a(X)$ to be $1$ if any two distinct polymers 
$\gamma_1,\gamma_2\in X$ are compatible, 
and $0$ if some distinct $\gamma_1,\gamma_2\in X$ has 
$\gamma_1\not\sim\gamma_2$.
We assume that a complex valued function $\phi(\gamma)$, 
$\gamma\in{\cal P}$ is given, and 
call $\phi(\gamma)$ the weight (or activity) 
of the polymer $\gamma$. 

Let a finite set $V\subset{\cal P}$ be given. 
The ensemble of all subsets $X\subset V$  
for which $a(X)=1$  will be called a ``gas'' of polymers,
while
$$
Z(V) = \sum_{X\subset V}
a(X)Ê\prod_{\gamma\in X} \phi(\gamma)
\eqno(A.1)$$
will be called the partition function in the volume $V$
of this gas 
(for $X=\emptyset$ the product in the above
expression is interpreted as the number 1).

We introduce the notion of 
multi-index as the maps $X$ from ${\cal P}$ into the set of
non negative integers, such that 
$$
N(X)=\sum_{\gamma\in{\cal P}} X(\gamma)<\infty
\eqno(A.2)$$
and define
$$
\phi^X=\prod_{\gamma\in{\cal P}} \phi(\gamma)^{X(\gamma)},
\quad
X!=\prod_{\gamma\in{\cal E}} X(\gamma)!
\eqno(A.3)$$
We denote by $\supp X$
the set of polymers for which $X(\gamma)\ne0$.
A multi-index can be viewed as a finite family of elements
of ${\cal P}$, $X(\gamma)$ being the multiplicity of
the element $\gamma$ in the family $X$.

To any multi-index one associates a graph
$G(X)$ with $N(X)$ points:
$X(\gamma_1)$ distinct vertices associated to $\gamma_1$,
etc.. 
We put a line between $\gamma_1,\gamma_2$, if, 
and only if,
$\gamma_1$ and $\gamma_2$ are not compatible
(in particular if $\gamma_1=\gamma_2$ we put
a line between them).
A subgraph $C$ of $G(X)$ is called a full subgraph if
the sets of vertices of $C$ and of $G(X)$ coincide.

Take $a^{\rm T}(X)=0$ if $G(X)$ is disconnected, and
$$
a^{\rm T}(X) = (X!)^{-1} \sum_{{\scriptstyle 
C\subset G(X)}\atop
{\scriptstyle C\  {\rm connected}}} (-1)^{\ell(X)}
\eqno(A.5)$$
otherwise. Here the sum runs over all connected full subgraphs 
of $G(X)$ and $\ell(X)$ is the number of lines of $C$
(if G(x) has a single point we interpret the sum as having
one graph with $\ell(C)=1$). 

\medskip

{\bf Theorem A.1.}
{\it We have}
$$
\ln Z(V) = \sum_{X:\ \supp X\subset V} a^{\rm T}(X)\ \phi^X 
\eqno(A.4)$$

\medskip

The connected $X$ will be called clusters (of polymers).
The expansions in terms of these objects are the cluster 
expansions. 
Their convergence properties will be 
established in the Theorem below. 
If the weights $\phi(\gamma)$ are small enough for ``large'' 
polymers $\gamma$, this Theorem   
enables one to make good evaluations of the free energy 
as well as of the decay of correlations. 
Let us remark also that whenever $\phi$ is analytic in some
parameter the estimates stated below imply the analyticity 
of the free energy and the correlation functions.

\medskip

{\bf Theorem A.2.}
{\it Assume that there exists a positive real valued function 
$\mu(\gamma)$, $\gamma\in{\cal P}$, such that
$$
\vert\phi(\gamma)\vert\ \mu(\gamma)^{-1}
\exp\Big(\sum_{\gamma'\not\sim\gamma}\mu(\gamma')\Big) \le r < 1
\eqno(A.6)$$
for each $\gamma\in{\cal P}$. Then the following estimate
$$
\sum_{X:\ \gamma\in\supp X} \vert a^{\rm T}(X)\ 
\phi^X\vert 
\le\mu(\gamma)r(1-r)^{-1}
\eqno(A.7)$$
holds true for every $\gamma\in{\cal P}$.}

\medskip

See Gallavotti et al. (1973), chapter 4, 
for the proof of the results stated in 
Theorems A.1 and A.2,  
in the particular case of the Ising model at low temperatures. 
In this case the term contour is used instead of polymer. 
To stress that the formulation, as well as the proof of these
results, do not depend on details of the ``geometry'' 
of polymers, we have presented them in an abstract setting.

\vfill\eject

\chapter
{\ttt 2.3\ \  Appendix B : Proof of Theorem 2.1 }  
\echapter

In this appendix we prove Theorem 2.1 for the 
three-dimensional Ising model. We write
$$
F(L_1,L_2,L_3) = F(\Lambda) = -(1/\beta)\ln\big(
Z^{(\pm,{\bf n})}(\Lambda)/Z^{+}(\Lambda)\big)
\eqno(B.1)$$
First, we establish the monotonicity 
property
$$
F(L_1,L_2,L_3)\leq 
F(L_1,L_2,L'_3)\quad 
\hbox{if}\quad  L_3\ge L_3'
\eqno(B.2)$$
for $L_d'$ large enough. 
This shows that the limit $L_3\to\infty$ exists and
$$
\lim_{L_3\to\infty}\ F(L_1,L_2,L_3)=\inf_{L_3}\
F(L_1,L_2,L_3)\equiv F(L_1,L_2) 
\eqno (B.3)$$
We shall then prove that 
$F(L_1,L_2)$ 
is bounded
$$0\le F(L_1,L_2)\le K_1 L_1 L_2
\eqno(B.4)$$
and that it satisfies the following subadditivity property 
(up to boundary terms)
$$
F(L_1'+L_1''+1,L_2)\le F(L_1',L_2) + F(L_1'',L_2)
+ K_2(L_1+L_2)
\eqno(B.5)$$
The subadditivity (B.5) together with the bound (B.4)
imply, following standard arguments in the theory of 
thermodynamic limits (see for instance Ruelle 1969) 
the existence of
$$\lim_{L_1,L_2\to\infty}\,
{1\over {L_1 L_2}}\, F(L_1,L_2)
\eqno(B.6)$$
Moreover this limit equals the infimum over $L_1,L_2$. 
This ends the proof of the first part of Theorem 2.1, 
provided properties (B.2), (B.4) and (B.5) are 
satisfied.

The proof of these properties 
uses Griffiths correlation inequalities 
(see for instance Ruelle 1969),
and is based in the following Lemma. 
This same Lemma will be used to prove the second part of 
Theorem 2.1.

\medskip

{\bf Lemma. }
Given two boxes $\Lambda,\Lambda'$ such that 
$\Lambda'\subset\Lambda$ we let $S_1$ be the set of 
sites of $\Lambda\setminus\Lambda '$ 
which are above or on the the plane $p_{\bf n}$ and
$S_2$ the sites of $\Lambda\setminus\Lambda '$ 
which are below this plane.
Then, we have
$$
0\le F(\Lambda)\le F(\Lambda')+2\beta N(S_1,S_2)
\eqno(B.7)$$
where $N(S_1,S_2)$ is the number of bonds 
$\langle i,j\rangle$ with $i\in S_1$ and $j\in S_2$.

\medskip

The monotonicity property (B.2) is just a particular
case of the above inequality (B.7). 
If $\Lambda$ and $\Lambda'$ are the two boxes considered 
in (B.2) one has $N(S_1,S_2)=0$. 

The upper bound (B.4) follows also from inequality (B.7) 
by taking $\Lambda'=\emptyset$ and noticing that
in this case $N(S_1,S_2)\le3L_1L_2$, 
which gives $K=3\beta J$.

In order to prove the subadditivity property (B.5) 
we consider three parallelepipeds 
$\Lambda_1$, $\Lambda_2$ and $\Lambda$ 
of sides 
$(L'_1,L_2,L_3'), (L''_1,L_2,L_3'')$ 
and $(L'_1+L''_1+1,L_2,L_3)$ 
placed in such a way that 
$\Lambda_1\cup\Lambda_2\subset\Lambda$, 
the distance from $\Lambda_1$ to $\Lambda_2$ is 1, 
and the plane $p_{\bf n}$ which passes through the center 
of $\Lambda$ passes also through the center of 
$\Lambda_1$ and $\Lambda_2$, and defines the 
$(\pm)$ boundary condition for the three boxes. 
Now we apply inequality (B.7) with   
$\Lambda'=\Lambda_1\cup\Lambda_2$. 
We notice that $\Lambda\setminus\Lambda'$ reduces to the 
vertical plane separating $\Lambda_1$ and $\Lambda_2$
when the vertical dimensions $L'_3$, $L''_3$
and $L_3$, tend to infinity.  
Thus $N(S_1,S_2)\le2L_1$, and $F(\Lambda')$,
the first term in the right hand side
of (B.7), nearly coincides with the sum 
$F(L'_1,L_2) + F(L''_1,L_2)$ because the partition
function $Z^{\bar\sigma}(\Lambda')$ factorizes. 

Some error is made because from the construction the 
centers of $\Lambda_1$ and $\Lambda_2$ 
do not necessarily coincide with a site of the lattice, 
as it is assumed in the definition of $\tau({\bf n})$. 	
This error however is bounded by the third term in the left 
hand side of (B.5) as it may be seen by using 
appropriate planes parallel to $p_{\bf n}$
introduced to compensate the displacement of the center 
(by a distance less than one) and applying again the
above argument.

We next prove that $\tau({\bf n})$ satisfies the pyramidal
inequality.

Let $A_0,\dots,A_3$ be the vertices of the pyramid. 
Introduce their projections $A'_0,\dots,$ $A'_3$ 
on the horizontal plane. and assume that $A'_0$ 
falls in the interior of the triangle $A'_1,A'_2,A'_3$. 
Let $\cal L'$ be the projection of the lattice $\cal L$. 
We denote by $Q_0$ the set of sites of $\cal L'$ 
inside the triangle $A'_1,A'_2,A'_3$ and by $Q_1$ 
the set of sites of $\cal L'$ inside the triangle
$A'_0,A'_2,A'_3$ whose distance from the sides of this 
triangle is larger than 1. 
Similarly we define the sets $Q_2$ and $Q_3$ 
with respect to the triangles
$A'_0,A'_1,A'_3$ and $A'_0,A'_1,A'_2$. 
We introduce the surface which coincides with the plane
$A_1,A_2,A_3$ outside the triangle $A'_1,A'_2,A'_3$ 
and with the other three faces of the pyramid inside it. 
We apply now a modified version of the Lemma 
in which this surface replaces the plane $p_{\bf n}$. 
Assume that $Q$ is the projection of the box $\Lambda$ 
and define $\Lambda'$ as the set of sites in $\Lambda$
whose projections belong to $Q_1 \cup Q_2 \cup Q_3$.
Then the modified Lemma implies
$$
F(Q_0)\leq F(Q_1)+F(Q_2)+F(Q_3)
+\vert Q_0 \setminus(Q_1 \cup Q_2 \cup Q_3)\vert 
\ (K/\alpha)
\eqno(B.8)$$
From this relation, the pyramidal inequality follows by 
passing to the limit when the three triangles tend to 
infinity.

\medskip

{\it Proof of the Lemma.Ê} 
To prove inequalities (B.7) 
we notice that $F(\Lambda')$ may be obtained from 
$F(\Lambda)$ by adding the external fields
$$ 
\eqalignno{
- h \sum_{i\in S_1\cup S_2}\sigma(i) \quad
& \hbox{to}\quad H(\sigma_{\Lambda}\mid +) \cr 
- h\sum_{i\in S_1}\sigma (i) + h\sum_{i\in S_2}\sigma (i)
\quad 
& \hbox{to}\quad  H(\sigma_{\Lambda}\mid\pm)
&(B.9)\cr}
$$
and letting $h$ tend to infinity.
In fact after this limit the partition functions 
$Z^{\bar\sigma}(\Lambda)$ in (B.1) are replaced by
$Z^{\bar\sigma}(\Lambda')$ multiplied by 
the term 
$$
\exp\big(\sum_{\langle i,j\rangle\subset
S_1\cup S_2}\bar\sigma(i)
\bar\sigma(j)\big)
$$ 
This means that $F(\Lambda)$ becomes equal
to the right hand side of equation (B.7).
But, Griffiths inequalities tell us that
$$
\langle\sigma(i)\rangle^{+}-\vert\langle\sigma(i)
\rangle^{\pm}\vert\ge0
\eqno(B.10)$$
where $\langle\quad\rangle^{\bar{\sigma}}$ 
denotes the expectation values corresponding to the Gibbs 
measure $(Z^{\bar{\sigma}})^{-1}$ $\exp(-\beta 
H(\sigma\mid\bar{\sigma})$, 
and show that the derivative with respect to $h$ of 
the modified $F(\Lambda)$ is positive, and 
hence that it is an increasing function of $h$.
We get therefore the second inequality in (B.7).
The same argument implies that 
$Z^{+}(\Lambda)$ is greater than $Z^{\pm}(\Lambda)$ 
and gives the first inequality in (B.7).

\def\aa{\item{}\kern -12pt\rm}
\def\bb{\ninerm}
\def\cc{\it}
\def\dd{\ninerm}
\def\ee{\ninerm}
\def\ff{\it}
\def\th{\thinspace}

\chapter
{\ttt References }
\echapter

\eightrm

\aa Abraham, D.B. (1986):
\bb Surface structures and phase transitions. 
\ee In: Domb, C., Lebo\-witz, J.L., eds.:
\ff Critical Phenomena, 
\dd vol.\th 10, pp.\th 1--74.
    Academic Press, London.

\aa Andreev, A.F. (1981): 
\bb Faceting phase transitions of crystals. 
\cc Sov. Phys. JETP 
\dd {\bf 53}, 1063.

\aa Avron, J.E., van Beijeren, H., Shulman, L.S., 
    Zia, R.K.P. (1982):
\bb Roughening transition, surface tension and the
    equilibrium droplet shapes in a two-dimensional  
    Ising system.
\cc J. Phys. A: Mat. Gen.  
\dd {\bf 15}, L 81--86. 

\aa Balibar, S., Castaign, B. (1980): 
\bb Possible observation of the roughening transition in 
    helium.
\cc J. Physique Lett. 
\dd {\bf 41}, L 369--332. 

\aa Balibar, S., Castaign, B. (1985): 
\bb Helium: solid-liquid interfaces. 
\cc Surf. Sci. Rep. 
\dd {\bf 5}, 87--143. 

\aa Beijeren, H. van (1975):
\bb Interface sharpness in the Ising model. 
\cc Commun. Math. Phys. 
\dd {\bf 40}, 1--6. 

\aa Beijeren, H. van, Nolden, I. (1987):
\bb The roughening transition. 
\ee In: Schommers, W., von Blackenhagen, P., eds.:
\ff Topics in Current Physics, 
\dd vol.\th 43, pp.\th 259--300. 
    Springer, Berlin.

\aa Bricmont, J., El Mellouki, A., Fr\"ohlich, J. (1986):
\bb Random surfaces in Statistical Mechanics:
    Roughening, rounding, wetting,...
\cc J. Stat. Phys.
\dd {\bf 42}, 743--798. 

\aa Bricmont, J., Fontaine, J.R., Lebowitz, J.L. (1982): 
\bb Surface tension, percolation and roughening.
\cc J. Stat. Phys. 
\dd {\bf 29}, 193--203. 

\aa Bricmont, J., Lebowitz, J.L., Pfister, C.E., 
    Olivieri, E. (1979):
\bb Non-trans\-lation invariant Gibbs states with 
    coexisting phases I. 
\cc Commun. Math. Phys. 
\dd {\bf 66}, 1--20. 

\aa Bricmont, J., Lebowitz, J.L., Pfister, C.E. (1979):
\bb Non-translation invariant Gibbs states with 
    coexisting phases II. 
\cc Commun. Math. Phys. 
\dd {\bf 66}, 21--36. 

\aa Bricmont, J., Lebowitz, J.L., Pfister, C.E. (1979):
\bb Non-translation invariant Gibbs states with 
    coexisting phases III. 
\cc Commun. Math. Phys. 
\dd {\bf 69}, 267--291. 

\aa Bricmont, J., Lebowitz, J.L., Pfister, C.E. (1980):
\bb On the surface tension of lattice systems.
\cc Ann. Acad. Sci. New York 
\dd {\bf 337}, 214--223. 

\aa Burago, Yu.D., Zalgaller, V.A. (1988):
\cc Geometric inequalities. 
\dd Grundieh\-ren der math. Wissenschaften, 
    vol.\th 43. 
    Springer, Berlin. 

\aa Curie, P. (1885):
\bb Sur la formation des cristaux et sur les constantes  
    capillaires de leurs differentes faces.
\cc Bull. Soc. Fr. Mineral. 
\dd {\bf 8}, 145-150. 
    Reprinted in Schneer (1977).  

\aa DeConinck, J., Dunlop, F., Rivasseau, V. (1989): 
\bb On the microscopic validity of the Wulff construction 
    and of the generalized Young equation.
\cc Commun. Math. Phys.
\dd {\bf 121}, 401--419. 

\aa Dobrushin, R.L. (1972):
\bb Gibbs state describing the coexistence of phases
    for a three dimensional Ising model.
\cc Theory Probab. Appl. 
\dd {\bf 17}, 582--600.

\aa Dobrushin, R.L., Koteck\' y, R., Shlosman, S.B. (1992):
\cc The Wulff Construction: 
    a Global Shape from Local Interactions.
\dd American Mathematical Society, Providence.

\aa Dobrushin, R.L., Koteck\' y, R., Shlosman, S.B. (1993):
\bb A microscopic justification of the Wulff construction.
\cc J. Stat. Phys. 
\dd {\bf 72}, 1--14.  

\aa Dobrushin, R.L., Hryniv, O. (1996):
\bb Fluctuations of shapes of large areas under
    paths of random walks.
\cc Probab. Theory Relat. Fields
\dd {\bf 105}, 423--458. 

\aa Dobrushin, R.L., Shlosman, S.B. (1985):  
\bb The problem of translation invariance of gibbs
    fields at low temperatures.
\cc Sov. Sci. Revs., Math. Phys. C
\dd {\bf 5}, 53--185.

\aa Dobrushin, R.L., Shlosman, S.B. (1992):  
\bb Thermodynamic inequalities and the geometry of 
    the Wulff construction.
\ee In: Albeverio, S. et al., eds.: 
\ff Ideas and Methods in Mathematical Analysis, 
    Stochastics and Applications,
\dd Cambridge University Press, Cambridge.

\aa Fr\"ohlich, J., Spencer, T. (1981):
\bb The Kosterlitz-Thouless transition in two-dimensio\-nal 
    abelian spin systems and the Coulomb gas.
\cc Commun. Math. Phys. 
\dd {\bf 81}, 527--602. 

\aa Galgani, L., Manzoni, L., Scotti, A. (1971):
\bb Asymptotic equivalence of equilibrium ensembles
    of classical statistical mechanics. 
\cc J. Math. Phys. 
\dd {\bf 12}, 933--935, 

\aa Gallavotti, G. (1972a):
\bb The phase separation line in the two-dimensional  
    Ising model.
\cc Commun. Math. Phys. 
\dd {\bf 27}, 103--136. 

\aa Gallavotti, G. (1972b):
\bb Instabilities and phase transition in the Ising model. 
    A review.
\cc Riv. Nuov. Cim. 
\dd {\bf 2}, 133--169. 

\aa Gallavotti,  G., Martin-Lof, A., Miracle-Sole, S.
    (1973):
\bb Some problems connected with the description of
    coexisting phases at low temperatures in Ising models.
\ee In: Lenard, A., ed.: 
\ff Mathematical Methods in Statistical Mechanics,
\dd Lecture Notes in Physics, vol.\th 20, pp.\th 162--204.
    Springer, Berlin. 

\aa Gibbs, J.W. (1875):
\bb On the equilibrium of heterogeneous substances,
\hfill\break
\cc Trans. Connenicut Acad. Arts Sci.
\dd {\bf 3}, 108--248 (1875) and 343--524 (1978).
    Reprinted in 
\cc The scientific papers of J.W. Gibbs, 
\dd vol.\th 1., pp.\th 55--353.
    Longmans, Green \& Co., London 1906.

\aa Goldschmidt, V. (1913--1923):
\cc Atlas der Kristallformen. 
\dd Universit\"ats\-verlag, Heidelberg. 

\aa Herring, C. (1951):
\bb Some theorems on the free energy of a crystal surface, 
\cc Phys. Rev.
\dd {\bf 82}, 87--93.

\aa Heyraud, J.C., M\'etois, J.J. (1980): 
\bb Establishment of the equilibrium shape of metal
    crystallites on a foreign substrate: gold on graphite, 
\cc Jour. Cryst. Growth 
\dd {\bf 50}, 571--574. 

\aa Heyraud, J.C., M\'etois, J.J. (1983): 
\bb Equilibrium shape and temperature: lead on graphite, 
\cc Surf. Sci. 
\dd {\bf 128}, 334--342. 

\aa Holick\'y, P., Koteck\'y, R., Zahradn\'\i k, M. (1988):
\bb Rigid interfaces for lattice models at low temperature.
\cc J. Stat. Phys. 
\dd {\bf 50}, 755--812. 

\aa Ioffe, D. (1994):
\bb Large deviations for the 2D Ising model:
    A lower bound without cluster expansions,
\cc J. Stat. Phys.
\dd {\bf 74}, 411--432.

\aa Ioffe, D. (1995):
\bb Exact large deviations bounds up to $T_c$
    for the Ising model in two dimensions,
\cc Probab. Theory Relat. Fields
\dd {\bf 102}, 313--330.

\aa Keshishev, K.O., Parshin, A.Ya., Babkin, A.V. (1981): 
\bb Crystallization waves in He$^4$. 
\cc Sov. Phys. JETP 
\dd {\bf 53}, 362--369. 

\aa Kotecky, R. (1989):
\bb Statistical mechanics of interfaces and equilibrium 
    crystal shapes. 
\ee In: Simon, B. et al., eds.: 
\ff IX International Congress of Mathematical Physics,
\dd pp.\th 148--163,
    Adam Hilger, Bristol.

\aa Kotecky, R., Miracle-Sole, S. (1986):
\bb A model with roughening transition at low temperatures. 
\cc Phys. Rev. B 
\dd {\bf 34}, 2049-2051. 

\aa Kotecky, R., Miracle-Sole, S. (1987a):
\bb Roughening transition for the Ising model on a bcc 
    lattice. A case in the theory of ground states.
\cc J. Stat. Phys. 
\dd {\bf 47}, 773--799. 

\aa Kotecky, R., Miracle-Sole, S. (1987b):
\bb A roughening transition indicated by the behavior 
    of ground states.
\ee In: Mebkhout, M., Seneor, R., eds.:
\ff VIIIth International Congress on Mathematical Physics,
\dd pp.\th 331--337,
    World Scientific, Singapore. 

\aa Kotecky, R., Pfister, C.E. (1994):
\bb Equilibrium shapes of crystals attached to walls. 
\cc J. Stat. Phys. 
\dd {\bf 76}, 419--445. 

\aa Laue, M. von (1943):
\bb Der Wulffsche Satz f\"ur die Gleichgewichtsform 
    von Kristal\-len.
\cc Z. Kristallogr.
\dd {\bf 105}, 124--133. 
    Reprinted in Schneer (1977).  

\aa Lebowitz, J.L., Pfister, C.E. (1981):
\bb Surface tension and phase coexistence.
\cc Phys. Rev. Lett. 
\dd {\bf 46}, 1031--1033. 

\aa Messager, A., Miracle-Sole, S., Ruiz, S. (1992): 
\bb Convexity properties of the surface tension and 
    equilibrium crystals.
\cc J. Stat. Phys.
\dd {\bf 67}, 449--470.  

\aa Miracle-Sole, S., (1976): 
\bb Theorems on phase transitions with a treatment for 
    the Ising model. 
\ee In: Brey, J. and Jones, R.B., eds.:
\ff Critical phenomena. 
\dd Lecture notes in Physics, vol.\th 80, pp.\th 190-215.
    Springer, Berlin. 

\aa Miracle-Sole, S., (1995a): 
\bb Surface tension, step free energy and facets in the
    equilibrium crystal shape. 
\cc J. Stat. Phys. 
\dd {\bf 79}, 183--214. 

\aa Miracle-Sole, S., (1995b): 
\bb On the microscopic theory of phase coexistence.
\ee In: Brey, J.J. et al., eds.:
\ff 25 Years of Non-Equilibrium Statistical Mechanics,
\dd pp.\th 312--321.
    Springer, Lecture Notes in Physics 445, Berlin.                   

\aa Miracle-Sole, S., Ruiz, J. (1994):
\bb On the Wulff construction as a problem of equivalence 
    of ensembles.
\ee In: Fannes, M. et al., eds.: 
\ff On Three Levels: Micro, Meso and Macroscopic Approaches 
    in Physics, 
\dd pp.\th 295--302.
    Plenum Press, New York.

\aa Pavlovska, A., Nenov, D. (1971):
\bb Experimental investigation of the surface melting of 
    equilibrium form faces of diphenil.   
\cc Surf. Sci.  
\dd {\bf 27}, 211--217. 

\aa Pavlovska, A., Nenov, D. (1971):
\bb Les surfaces non-singuli\`eres sur la forme 
    d'\'equilibre du naphtal\`ene. 
\cc Jour. Cryst. Growth  
\dd {\bf 12}, 9--12. 

\aa Pavlovska, A., Nenov, D. (1977):
\bb Experimental study of the surface melting of 
    tetra\-brom\-methane.  
\cc Jour. Cryst. Growth  
\dd {\bf 39}, 346--352. 

\aa Pfister, C.E. (1991):
\bb Large deviations and phase separation in the 
    two-di\-men\-sio\-nal Ising model.
\cc Helv. Phys. Acta 
\dd {\bf 64}, 953--1054. 

\aa Rottman, C., Wortis, M. (1984):
\bb Statistical Mechanics of equilibrium crystal shapes:
    Interfacial phase diagrams and phase transitions.
\cc Phys. Rep. 
\dd {\bf 103}, 59--79. 

\aa Ruelle, D. (1969): 
\cc Statistical Mechanics, Rigorous Results. 
\dd Benjamin, New York. 

\aa Schneer, C.J., ed., (1970):
\bb Morphological basis for the reticular hypothesis 
\cc Am. Mineralogist 
\dd {\bf 55}, 1466--1488.
    Reprinted in Schneer (1977). 

\aa Schneer, C.J., ed., (1977):
\cc Crystal Form and Structure. 
\dd Benchmark papers in Geology, vol.\th 34. 
    Dowden, Hutchinson \& Ross, Stroudsbourg (Penn).

\aa Taylor, J.E. (1987): 
\bb Some crystalline variational techniques and results. 
\cc Asterisque
\dd {\bf 154-155}, 307--320.  

\aa Weeks, J.D., Gilmer, G.H., Leamy, H.J. (1973): 
\bb Structural transition in the Ising model interface.
\cc Phys. Rev. Lett.
\dd {\bf 31}, 549--551.   

\aa Winterbottom, W.L. (1967): 
\bb Equilibrium shape of a small particle in contact
    with a foreign substrate. 
\cc Acta Metallurgica
\dd {\bf 15}, 303--310. 

\aa Wolf, P.E., Balibar, S., Gallet, F. (1983): 
\bb Experimental observations of a third roughening 
    transition in hcp $^4$He crystals. 
\cc Phys. Rev. Lett. 
\dd {\bf 51}, 1366--1369. 

\aa Wulff, G. (1901):
\bb Zur Frage der Geschwindigheit des Wachstums und der 
    Auflosung der Kristallflachen. 
\cc Z. Kristallog. 
\dd {\bf 34}, 449--530. 
    Reprinted in Schneer (1977).  

\end